\documentclass[a4paper,fleqn,usenatbib,useAMS]{mnras}

\usepackage{graphicx}	
\usepackage{amsmath}	
\usepackage{amssymb}	
\usepackage{multicol}        
\usepackage{bm}		
\usepackage{pdflscape}	
\usepackage{journals}
\usepackage{epstopdf}

\newcommand{\Msun}{M\ensuremath{_\odot}\,}

\def\HI{H\,{\sc i}\, }

\def\kms{$\textrm{km~s$^{-1}$}$}
\def\nb{\textsc{nbursts}}

\usepackage[T1]{fontenc}
\usepackage{ae,aecompl}

\usepackage{newtxtext,newtxmath}
\usepackage{epstopdf}

\title[A study of UGC~1378 -- galaxy with giant disc ]{UGC~1378 -- a Milky Way-sized galaxy embedded in a giant low-surface brightness disc}
\author[A. Saburova et al.]{
Anna S. Saburova,$^{1,2}$ \thanks{E-mail:saburovaann@gmail.com}
Igor V. Chilingarian,$^{3,1}$
Anastasia V. Kasparova,$^{1}$
Ivan Yu. Katkov,$^{1, 4}$ 
\newauthor
Daniel G. Fabricant,$^{3}$
Roman I. Uklein,$^{5}$
\\
$^1$ Sternberg Astronomical Institute, Moscow M.V. Lomonosov State University, Universitetskij pr., 13,  Moscow, 119234, Russia\\
$^2$ Institute of Astronomy, Russian Academy of Sciences, Pyatnitskaya st., 48, 119017 Moscow, Russia\\
$^3$ Smithsonian Astrophysical Observatory, 60 Garden Street MS09, Cambridge, MA 02138, USA\\
$^4$ New York University Abu Dhabi, Saadiyat Island, PO Box 129188, Abu Dhabi, UAE\\
$^5$ Special Astrophysical Observatory, Russian Academy of Sciences, Nizhniy Arkhyz, Karachai-Cherkessian Republic 357147, Russia\\
}

\begin{document}
\label{firstpage}
\pagerange{\pageref{firstpage}--\pageref{lastpage}} \pubyear{2018}
\maketitle

\begin{abstract}
The dominant physical processes responsible for the formation and longevity of giant gaseous and stellar discs in galaxies remain controversial. Although they are rare (less than 10 confirmed as of now), giant low-surface brightness (gLSB) discy galaxies provide interesting insights given their extreme nature.  We describe observations of UGC~1378 including deep spectroscopy with the Russian 6m telescope and multi-band imaging with Binospec at the MMT.  Galaxy UGC~1378 has both high surface brightness and an extended low surface brightness discs. Our stellar velocity dispersion data for the high-surface brightness, Milky Way-sized, disc appears inconsistent with a recent major merger, a widely discussed formation scenario for the very extended low surface brightness disc. We estimate the star formation rates (SFR) from archival Wide-Field Infrared Survey Explorer data. The SFR surface density in the LSB disc is low relative to its gas density, consistent with recent gas accretion. We argue that the unusually large size of UGC~1378's disc  may be the product of a rich gas reservoir (e.g. a cosmic filament) and an isolated environment that has preserved the giant disc.  
\end{abstract}

\section{Introduction}\label{intro}

We report observations of an unusual massive gas-rich galaxy, UGC~1378, with extended (radii $\sim 50$~kpc) gaseous and low-surface brightness stellar discs \citep{Mishra2017}. We build upon our previous studies of giant low-surface brightness galaxies (gLSB) in \citet{Kasparova2014} and \citet{Saburovaetal2018}. These galaxies deserve special attention because the formation of a $\sim10^{12}$~M\sun~disc galaxy is not easily explained in the hierarchical clustering paradigm. Major merger events are expected to destroy large discs but in simulations the merger trees of $10^{12}$~M\sun ~haloes almost always include major mergers \citep{Rodriguez-Gomez2015}.  We would like to understand how these large discs form, how they managed to survive, and whether there is a continuum of properties between gLSB and HSB galaxies. Studies of gLSB galaxies can also help us to understand the importance of gas accretion and outflows. 

UGC~1378 is classified as a type-SBa gLSB system \citep{Schombert1998} and is especially intriguing because its \HI content is higher than expected relative to its stellar mass \citep[according to the relation in][]{Wang2013}. Our deep imaging reveals the complex structure of UGC~1378, including a high-surface brightness (HSB) bulge and disc, and a gLSB disc. In common with other gLSBGs, UGC~1378 has prominent spiral arms \citep[like Malin~1, Malin~2, UGC~1922, UGC~1382,][]{Boissier2016, Galaz2015,Hagen2016, Kasparova2014, Saburova2018}, resembling the Milky Way immersed in a gLSB disc.  

Formation scenarios for gLSBs include: (i) a major merger  \citep[][]{Zhu2018, Saburovaetal2018} in which an extended disc is formed from an ample supply of gas cooled down at the late stage of a merger; (ii) a  gradual build-up of a gLSB disc by minor mergers with gas-rich satellites \citep{2006ApJ650L33P}; (iii) a build-up of a gLSB disc by accretion from  cosmic filaments \citep{Saburovaetal2018}; (iv) a non-catastrophic scenario in which the peculiar structure of a gLSB is created because of the large radial scale and low central density of a dark halo \citep{Kasparova2014}.

Accretion of intergalactic gas is required to explain star formation rates in disc galaxies \citep{Larson1980}  and also affects the angular momentum of galactic discs \citep[see, e.g.][and references therein]{Stewart2017}. The kinematics of extended disc-like structures traced by cold gas around galaxies at $z\approx 1$ is consistent with a cold accretion model \citep{Zabl2019}.  There is meagre direct observational evidence for accretion  in massive galaxies at low redshifts \citep[see, e.g.][]{Oosterloo2007, Vollmer2016, Fraternali2001, Fraternali2002, deBlok2014} even though in low-mass galaxies it is more commonly observed \citep[see, e.g.][]{Kirbyetal2016, Schmidt2014, J2007}. At the same time, the radial distribution of the oxygen effective yield in massive spirals indicates that accretion of metal poor-gas must be higher in larger galaxies \citep{Zasovetal2015}. Searches for accretion in massive \HI excess galaxies may therefore be productive. 

\HI excess galaxies have been the focus of previous studies. The ``Bluedisk'' project \citep{Wang2013} mapped the \HI distribution in 23 nearby galaxies with unusually high \HI mass fractions, finding an \HI excess in the surrounding environment \citep{Wang2015}. Kinematical studies of 'Bluedisk' galaxies did not reveal enhanced global asymmetry of the H\,{\sc i}-excess galaxies relative to the control sample, a possible indication of accretion \citep{denHeijer2015}.  A study of five \HI excess galaxies with inefficient star formation and stellar masses $>10^{10}$~\Msun gas by \citet{Gereb2018} suggested an external gas origin in two cases and an ambiguous gas origin in the other cases.  UGC~1378 is more massive than most of the ``Bluedisk" galaxies and has a redshift roughly half the ``Bluedisk'' mean. UGC~1378's gas-to-stellar mass ratio, its \HI disc size and mass are similar to the ``Bluedisk'' galaxies.  UGC~1378's \HI mass fraction is lower than that of the galaxies in the \citet{Gereb2018} and HIghMass \citep{highmass} samples. UGC~1378's properties are summarized in Table \ref{properties}.

Like most other gLSBs, UGC~1378 is located in a low-density environment \citep{Saburovaetal2018}, therefore the dynamical disturbances from a dense environment are minimized and it is easier to search for cold accretion. The \citet{Saulder2016} group catalogue contains no spectroscopically confirmed companions for UGC~1378 and our deep photometry reveals no satellites with {\it g} surface brightness higher than 26~mag~arcsec$^{-2}$ within 80~kpc.  The sparse environment may be the key factor to the survival of a large low-surface brightness disc even though it does not necessarily mean that the galaxy grew in isolation throughout the cosmic time. UGC~1378 may have accreted the absent satellites.  However, the accretion did not transform UGC~1378 into an early-type galaxy \citep[see, e.g.][]{Deeley2017} and it remains a discy system. We suggest that UGC~1378 may have formed in two stages. The first stage, in common with MW-type galaxies, may have included several episodes of merging and a second stage, following the accretion of most satellites, quiescently formed gLSB stellar and gaseous discs by accretion of metal-poor gas from a cosmic filament.

This paper is organized as follows: Section~\ref{Obs} describes long-slit spectral and photometric observations and data reduction; the results of the data analysis are given in Section~\ref{Res}; our mass modelling, which estimates the masses of the discs, bulge and dark matter halo is described in Section~\ref{massmod}; we discuss the results in Section~\ref{Discussion}; and summarize our findings in Section~\ref{conclusion}.

\begin{table}
\caption{Basic properties of UGC~1378. \newline 
References: {[1]} NED (http://ned.ipac.caltech.edu), {[2]} \citet{Mishra2017}. \label{properties}
\label{tab1}}
 \begin{center}
 \begin{tabular}{lll}
\hline \hline
Names& UGC~1378&ref.\\
        & PGC~007247 &\\
        & CGCG326-002&\\
\hline
Equatorial coordinates  & 01h56m19.2s &\\ 
(J2000.0)& +73d16m58s&[1]\\
Distance     & $38.8$ Mpc& [2]\\
Morphological type     & 	SBa&[1]\\
Inclination angle     & $59$\degr&[2]\\
Major axis position angle    &$181$\degr&[2]\\
\HI mass &$1.2 \times 10^{10}$\Msun&[2]\\
Rotation velocity& 282 \kms&[2]\\
Scale kpc/arcsec&0.188&\\
\hline
\end{tabular}

\end{center}
\end{table}

\section{Observations and data reduction}\label{Obs}
\subsection{Long-slit spectral observations}
We observed UGC~1378 with the SCORPIO focal reducer/spectrograph  \citep{AfanasievMoiseev2005} at the prime focus of the 6-m Russian telescope on 20 September 2017 using a 6-arcmin long and 1~arcsec wide slit, and the VPH G2300G grism. This setup provides spectral coverage from 4800-5570 \AA, dispersion of 0.38 \AA\ pixel$^{-1}$, with an instrumental FWHM of 2.2 \AA. The plate scale along the slit is 0.36~arcsec pixel$^{-1}$. The total on-source exposure time  was 7 hours in 1.4~arcsec seeing. The slit was oriented along the major axis of UGC~1378 at $PA=181$\degr \citep{Mishra2017}. We used spectra of on-board arc and flat field lamps for calibration.

We reduced the spectral data using our \textsc{idl} based pipeline. The data reduction steps include: bias subtraction and overscan clipping, flat-field correction, wavelength calibration\footnote{To improve the wavelength solution accuracy we took arc spectra every 2~h and used them to reduce the corresponding science frames.}, cosmic ray removal, linearization, and co-addition of multiple exposures. Night sky subtraction used the algorithm described in \citet{KatkovChilingarian2011}  and flux calibration used observations of the spectrophotometric standard star \emph{BD~25+4655}.

We measured the instrumental line-spread function of the spectrograph along the slit throughout the observed wavelength range, by fitting a twilight sky spectrum (observed in the same night) to a high-resolution Solar spectrum using {\sc ppxf} \citep{CE04}. We then fitted the reduced spectra of UGC~1378 with intermediate-resolution ($R=10000$) PEGASE.HR~\citep{LeBorgneetal2004} simple stellar population models (SSP) for Salpeter IMF \citep{Salpeter1955} convolved with the instrumental line-spread function using the \nb{} full spectral fitting technique  \citep{Chilingarian2007a, Chilingarian2007b}.  The fitting procedure returns the best-fitting parameters of an SSP model, that is age T~(Gyr) and metallicity [Fe/H]~(dex) of stellar population. We parametrized the line-of-sight velocity distribution (LOSVD) of stars by Gauss-Hermite series  \citep[see][]{vanderMarel1993}.  We obtained a luminosity-weighted stellar age and metallicity, line-of-sight velocity, velocity dispersion and Gauss-Hermite moments $h_3$ and $h_4$ which characterize the deviation of LOSVD from the Gaussian profile. We set $h_3$ and $h_4$ as zero starting from $\pm 15$ arcsec, since we were not able to define reliably these parameters at large galactocentric distances. We defined spatial bins along the slit for the fitting manually by first running an automatic 1D adaptive binning procedure and then adjusting the bin sizes to handle the sharp change in the surface brightness along the slit.
We separately analyzed the emission spectrum obtained by subtracting the best-fit stellar populations from the observed spectra. We fitted emission lines to a single Gaussian profile and derived the velocity and velocity dispersion of the ionised gas.\footnote{The results of the spectral data analysis in txt-format are available via https://doi.org/10.5281/zenodo.3352306 }

\subsection{Photometric observations}
\begin{figure}
\includegraphics[width=1\linewidth]{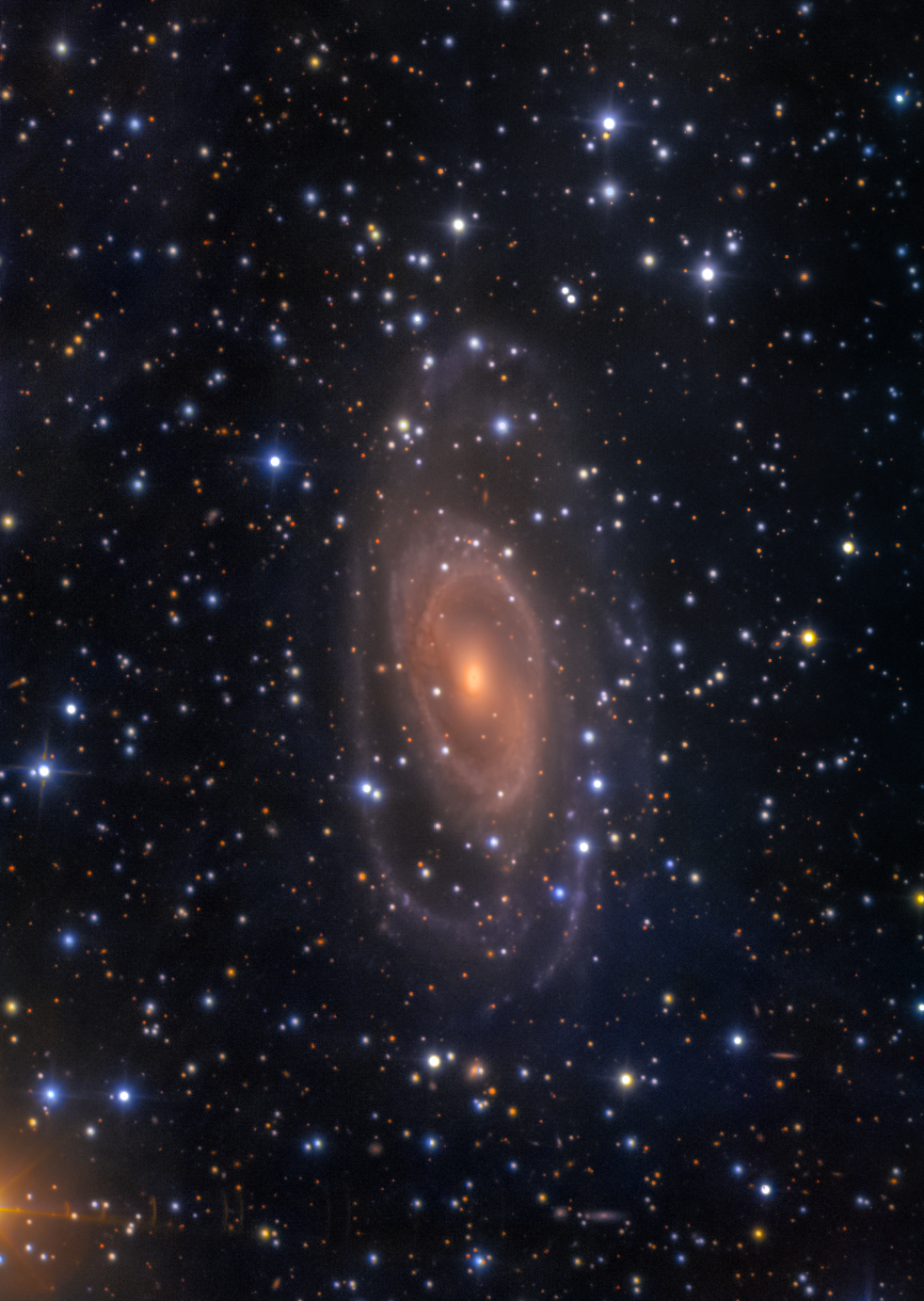}
\caption{Composite {\it r, g, z}-band image of UGC~1378. North is up, east is left, the image size is 7.2$\times$10.1~arcmin.}
\label{image}
\end{figure}

We observed UGC~1378 in the {\it g,r,z} bands with the Binospec spectrograph/imager \citep{Fabricant2019} mounted at the 6.5-meter converted MMT at Mt.Hopkins, Arizona during the Binospec commissioning run on 15 November 2017. The images were reduced using the Binospec pipeline \citep{Kansky2019}. The reduction included cosmic ray cleaning, flat-fielding, alignment of images, illumination correction, co-adding of image stacks, astrometric calibration using {\sc scamp} and re-mapping to the tangential projection using {\sc swarp} \citep{swarp}. After that we subtracted the sky with an \textsc{idl} code. We performed the aperture photometry of stars visible in the images and matched them against the sources in the Data Release 1 of the Pan-STARRS1 Survey \citep[PS1,][]{Chambers2016_panstarrs1}. Fig. \ref{image} is our composite {\it g,r,z} image of UGC~1378.

\section{Data analysis results}\label{Res} 

\subsection{Radial profiles of kinematics and stellar populations}

\begin{figure}
\includegraphics[width=1\linewidth]{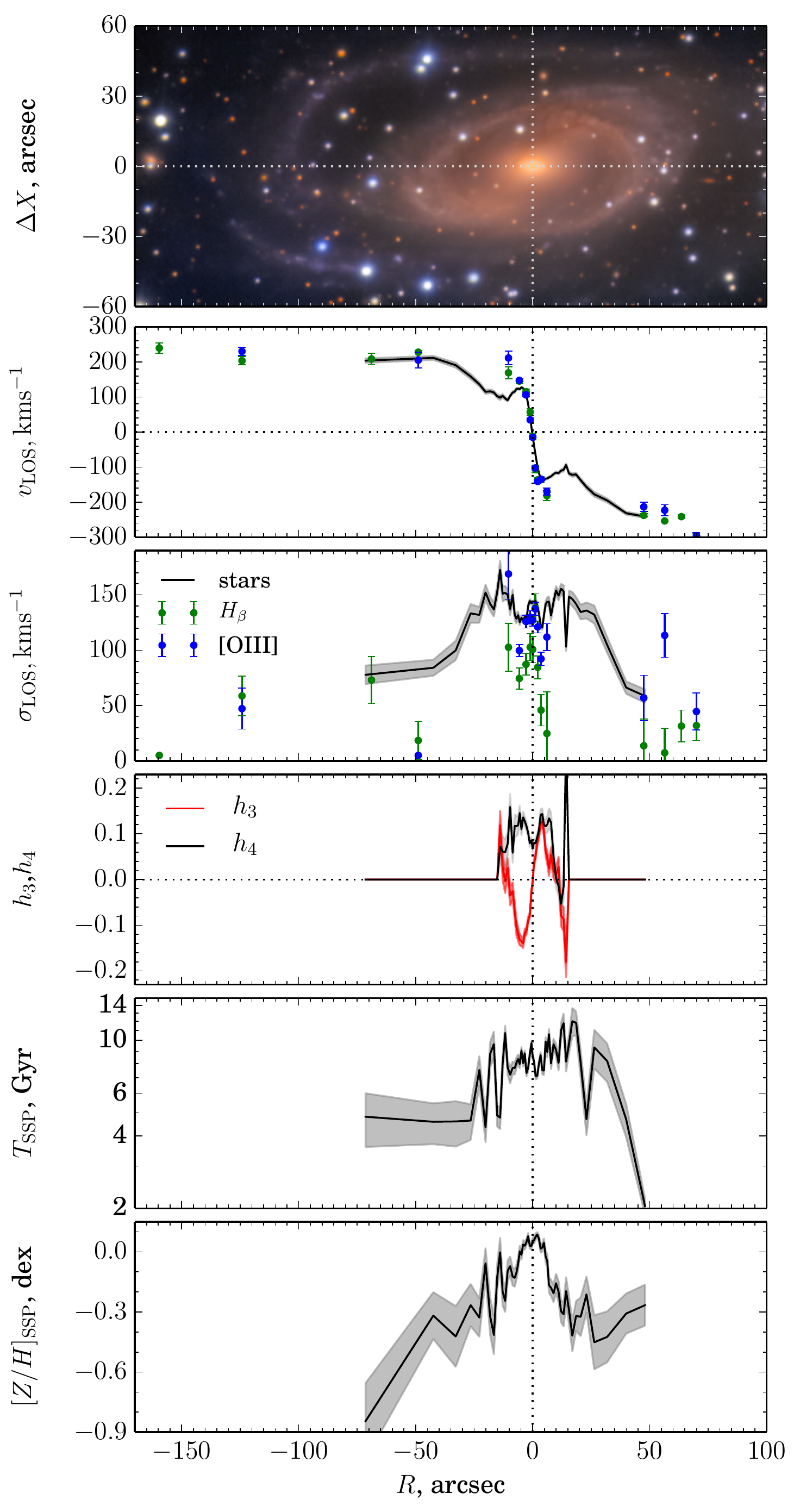}
\caption{The radial profiles of kinematics and stellar populations of UGC~1378  ($PA=181$\degr). From top to bottom: the slit position overlaid on a composite \textit{grz}-image; LOS velocity of ionized gas measured in $H_\beta$ and [O{\sc iii}] lines (circles) and stars (solid line and shaded areas are for values and their uncertainties),  the adopted systemic velocity is 2930 \kms; LOS velocity dispersion of gas (circles) and stars (line); $h_3$ and $h_4$ Gauss-Hermite coefficients; SSP-equivalent age of stellar population; SSP-equivalent metallicity of stellar population.}
\label{profiles_sp}
\end{figure}

Fig.~\ref{profiles_sp} summarizes our spatially resolved spectral results.  The top panel displays the orientation of the long slit along UGC~1378's major axis.  The second panel from the top displays the line-of-sight gas and stellar velocities along the slit, while the third panel plots the gas and stellar velocity dispersions. The strong velocity gradient in the stellar velocity profile indicates a strong bar \citep[see, e.g.][]{Saburova2017} that is also apparent in the image. The presence of the bar is also evident in the fourth panel in the radial profiles of stellar Gauss-Hermite moments $h_3$ and $h_4$ that characterize the LOSVD deviation from Gaussian. The $h_3$ profile anticorrelates with radial velocity in the inner region, typical of galaxies with boxy or peanut-shaped bulges \citep{Chung2004}. This behaviour of $h_3$ is also found in galaxies with an inner disc or a ring \citep[see, e.g.][]{Seidel2015}.  The central high velocity dispersion and old metal-rich stellar population (see, Fig. \ref{profiles_sp}, two bottom panels) are, however, inconsistent with a kinematically cold young stellar component in the inner part expected with a star-forming ring. The central minimum of the $h_4$ profile is also reproduced in numerical simulations of a barred galaxy \citep[][]{Saburova2017}. In the centre of UGC~1378 we see a (pseudo-)bulge (see below) dominated by old, metal-rich stars, formed due to the presence of the bar \citep{Kormendy2004}. 

\subsection{Light profile decomposition}

\begin{figure}
\vspace{-35.8cm}
\hspace{-1cm}
\includegraphics[width=7.3\linewidth]{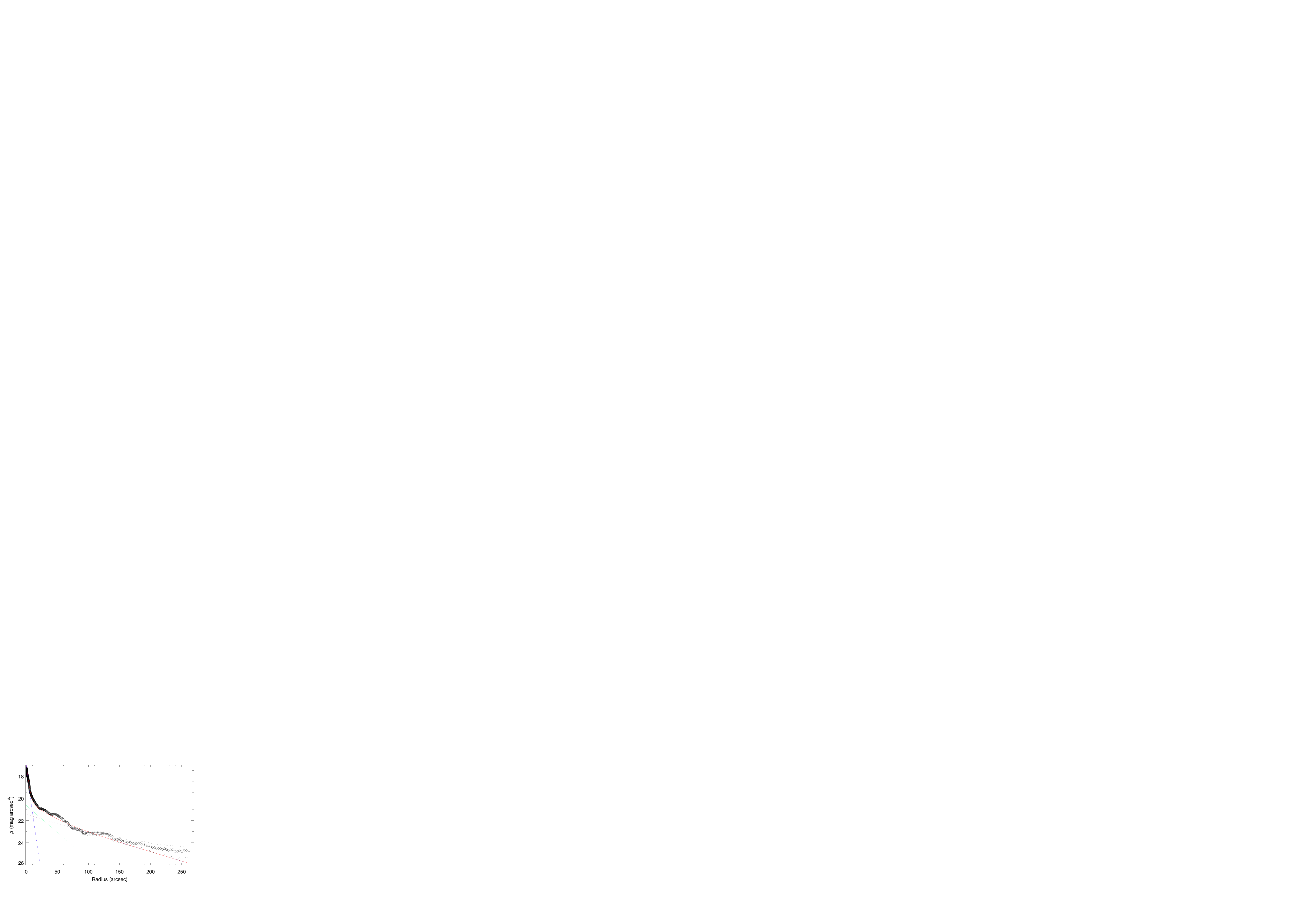}
\vspace{-2.2cm}
\caption{The {\it g}-band surface brightness profile of UGC~1378 with an overplotted best-fitting {\sc galfit} model. HSB and LSB discs are shown as green and brown lines, bulge as a blue line, and the total model as a red line.  The profiles are corrected for Galactic extinction and are not corrected for inclination.}
\label{profiles}
\end{figure}

We fitted two dimensional models to the deep {\it g,r,z}-band images.  UGC~1378 is located at 11$^{\circ}$ galactic latitude so that variable Galactic extinction could introduce systematic errors if we attempt one dimensional fits to the radial profile. We use GALFIT \citep{Peng2002} to fit three structural components, (i) a Sersic bulge \citep{Sersic68}, parametrized by the central surface brightness $(I_0)_b$, the effective (half-light) radius $R_e$, and  the Sersic index $n$; (ii) two exponential discs defined by  the central surface brightness $(I_{d})_{0}$ and the exponential scale length $R_d$. The observed profile is poorly described by a model with a single disc and Sersic bulge, which is evident from Fig. \ref{profiles} where we show the {\it g}-band azhimuthaly averaged profile of surface brightness\footnote{We  calculated a surface brightness profile from the {\it g}-band image  using the {\sc ellipse} routine \citep{Jedrzejewski1987} in the {\sc iraf} software environment (\citealt{iraf}).} which is in a good agreement with the results of 2D fitting with two discs. Thus we include a second disc. We masked foreground stars and the outer ring of the bar with $r=2$~kpc.  The derived parameters from the 2D decomposition are listed in Table \ref{phot_par}\footnote{We compared the results of 2D fitting with  that following from the decomposition of the surface brightness profiles using Levenberg-Marquardt non-linear least-squares \textsc{idl} routine \citep{Chilingarian2009} and found a good agreement which could evidence in favour of the reliability of our 2D modelling. The good agreement between 1D and 2D fittings is also evident from Fig.~\ref{profiles} where the 2D model does not deviate from the 1D profile except for the outer region which is effected by the Galactic cirrus.}.  The magnitudes in Table \ref{phot_par} are corrected for Galactic extinction but not for inclination. The error-bars obtained using GALFIT should be treated with caution since they could be significantly underestimated \citep[see, e.g.][]{Zhao2015}, in our 1D fit we obtained the error of 3-6 percent for the radial scales of HSB disc and bulge, 1 percent error for Sersic index, about 20 percent for the scale of LSB disc and up to  0.1 mag arcsec$^{-2}$ for the surface brightnesses which is more reasonable.  We note that the central surface brightness of the second disc appears to be higher than the usually adopted B-band LSB limit of $>$22 mag~arcsec$^{-2}$ (cf. the 21.65$\pm$0.3 range found by \citet{Freeman1970} in HSB galaxies; \citet{obc1997}). However our direct $g$-band estimate is affected by Galactic cirrus. The {\it r}-band measurement is more reliable, 21.54~mag~arcsec$^{-2}$, corresponding to 21.84 in the {\it g}-band using a $(g-r)$ colour of 0.3~mag for the LSB-disc (see the colour map below). We obtain a B-band central surface brightness $\mu_{0,B}=22.14$~mag~arcsec$^{-2}$ using transformations from \citet{Jester2005}. Hence, the second disc of UGC~1378 satisfies the LSB central surface brightness criterion despite being brighter than e.g. Malin~1's disc \citep{Boissier2016}.  In the {\it r} band, the radius of the LSB disc  \citep[4 disc radial scale lengths,][]{Kregel2004} is 268~arcsec or 50~kpc, which is similar to that of the \HI disc (45~kpc). The radius of the HSB disc $4(R_d)_r\approx 18$~kpc, which is roughly half of the \HI disc radius. The  bulge Sersic index is $\sim$1 indicating that UGC~1378 possesses a pseudo-bulge.

\begin{table*}
\caption{The photometric parameters of the two discs and the bulge of UGC~1378 estimated from a two-dimensional {\sc galfit} decomposition, brightnesses are corrected for extinction following \citep{Schlafly2011}  but not corrected for the inclination: central bulge surface brightness $(\mu_0)_b$; bulge effective radius $R_e$; bulge Sersic index $n$; HSB exponential disc central surface brightness $(\mu_0)_d$; HSB disc radial scale length $R_d$; LSB exponential disc central surface brightness $(\mu_0)_{d2}$; LSB disc radial scale length $R_{d2}$.  The scale is 0.188~kpc~arcsec$^{-1}$. \label{phot_par}}
\begin{center}
\begin{tabular}{cccccccc}
\hline\hline
Band&$(\mu_0)_b$&$R_e$&$n$&$(\mu_0)_d$&$R_d$&$(\mu_0)_{d2}$&$R_{d2}$\\
&(mag arcsec$^{-2}$)&(arcsec)&&(mag arcsec$^{-2}$)&(arcsec)&(mag arcsec$^{-2}$)&(arcsec)\\
\hline
{\it g} &$ 16.90\pm 0.00 $ & $4.30 \pm  0.01$ & $1.10\pm 0.01 $ & $20.64\pm0.00$ & $22.13\pm0.05 $& $21.44\pm 0.00 $&$64.16\pm0.09$ \\
{\it r} & $16.06\pm0.00 $ & $4.24\pm0.01 $ & $1.05\pm0.00 $ & $19.88\pm0.00 $& $24.00\pm0.03$&$21.54\pm0.00 $&$66.97\pm0.15$\\
{\it z} & $15.72\pm0.00 $ & $4.48\pm0.01 $ & $1.00\pm0.00 $ & $19.13\pm0.00 $& $22.34\pm0.03$&$21.04\pm0.01$&$63.92\pm0.18$\\
\hline
\end{tabular}
\end{center}
\end{table*}

\subsection{Mass-modelling of the rotation curve}\label{massmod}

We built UGC~1378's composite rotation curve using \HI \citep{Mishra2017} and ionized gas (this work) to measure the dynamical masses and sizes of its structural components.  For the optical data analysis we use the inclination value adopted by \citet{Mishra2017} for their \HI data. Our rotation curve, derived from emission lines of ionised gas, is in good agreement with the \HI data in the region where the two datasets overlap  (see Fig. \ref{massmod}, bottom panel, where we show \HI data and ionized gas with crosses and dots, respectively).  We decomposed the composite rotation curve into two stellar exponential discs (HSB and LSB), a Sersic bulge, a dark matter halo, and a gas disc\footnote{We calculated the \HI surface density from the 0th moment maps from \citet{Mishra2017} using {\sc ellipse} routine and took into account the expected contribution of helium (multiplied the \HI surface density by 1.3).}.  The weights of \HI and optical data points for the fitting procedure were taken from the error-bars shown in the bottom panel of Fig.~\ref{massmod}. The low S/N optical rotation curve data at $r>100$ arcsec were not used for mass modelling.

The rotation curves of the components were constructed as in \citet{Saburova2016}. We used the following dark halo density profiles: \\
(i) a Burkert density profile \citet{Burkert1995}:
\begin{equation}\label{Burkert}
\rho_{\mathrm{burk}}(r)=\frac{\rho_0 R_s^3}{(r+R_s)(r^2+R_s^2)}.
\end{equation}
Here $\rho_{0}$ and $R_{s}$ are the central density and the radial scale of the halo\footnote{Below $R_s$ and $\rho_0$ are different for the various DM density profiles.}.\\
(ii) a pseudo-isothermal profile (hereafter, piso):
\begin{equation}
\rho_{\mathrm{piso}}(r)=\frac{\rho_{0}}{(1+(r/R_{s})^2)}.
\end{equation}\\
(iii) a Navarro-Frenk-White profile \citet{nfw1996}  (hereafter, NFW): \begin{equation}
\rho_{\mathrm{nfw}}(r)=\frac{\rho_{0}}{(r/R_s)(1+(r/R_{s })^2)^{2}}.
\end{equation}

It is important to constrain the contributions of the stellar disc and bulge to the total mass because the rotation curve decomposition is ambiguous  \citep[see, e.g.][]{Saburova2016}.  We are able to limit the disc mass using the marginal gravitational stability criterion from stellar velocity dispersion measurements  beyond two  disc radial scale lengths from the centre   \citep[see, e.g.][]{Zasov2004, Saburova2011, Saburova2013}. The majority of late-type disc galaxies and some S0 galaxies have discs that are marginally gravitationally stable at two disc radial scale lengths \citep{Zasov2011}. 

We formulate the stability criterion as follows. A single component isothermal disc is locally marginally stable when the radial stellar velocity dispersion $\sigma_r$ at the distance from the centre $R$ is equal to the critical value:
\begin{equation}
(\sigma_{r})_{\rm crit}=Q_T\cdot 3.36G \Sigma_d / \varkappa,
\end{equation}
where $\varkappa$ is the epicyclic frequency, $\Sigma_d$ is the disc surface density  and $Q_T$ is Toomre's stability  parameter (equal to unity for pure radial perturbations of a thin disc). Numerical simulations of the marginal stability of finite thickness exponential discs show that the parameter  $Q_T\approx 1.2 - 3$ slowly increases with radius \citep[see e.g.][]{Khoperskov2003}.   The presence of a cold gaseous component can make a disc less stable \citep[see, e.g.][]{Romeo2011}, but the gas surface density in UGC~1378 at two HSB disc radial scalelengths is roughly 20 times lower than that of the stars, so this effect is negligible. The situation is complicated by the presence of the stellar LSB disc. According to the mass-modelling of the rotation curve (see below), its density may be only $2-3$ times lower than the HSB disc at this radius, so the LSB disc can influence the stability if its velocity dispersion differs significantly from that of HSB disc. Unfortunately, our data are not deep enough to trace the stellar velocity dispersion at the radii where the LSB disc dominates the surface brightness to evaluate its effect on the stability.  

We estimate the radial velocity dispersion from the observed line-of-sight stellar velocity dispersion $\sigma_{\rm obs}$ taking into account the expected links between the dispersion along the radial, azimuthal and vertical directions:

\begin{multline}
\sigma_{\rm obs}^2(r) = \sigma_z^2 \cdot \cos^2 (i)+ \sigma_{\phi}^2\cdot \sin^2 (i)\cdot \cos^2(\alpha) + \\ +\sigma_{r}\sin^2(i)\cdot \sin^2(\alpha),
\end{multline}
where $\alpha$  is the angle between the direction of the slit and the major axis.

To solve the equation we used two additional conditions: $ \sigma_{r} = 2\Omega
\cdot  \sigma_{\phi} /\varkappa $ (Lindblad formula for the epicyclic approximation) and $ \sigma_z =k\cdot \sigma_{r} $. The coefficient $k$  was taken to be 0.6 based on direct measurements that give an expected range 0.5--0.8 \citep[see e.g.][]{Shapiro2003}. We estimated the epicyclic frequency from the combined optical and \HI rotation curve \citep{Mishra2017} using the equation: $\varkappa(r)=2v(r)/r\sqrt{0.5+r/2v(r)( \frac{\partial v(r)}{\partial r})}$. We used the radial variation of the parameter  $Q_T$ resulting from the numerical simulations by \citet{Khoperskov2003}.

The central surface density of the HSB disc, extrapolated from an exponential model with the photometric scale length, and using the marginal gravitational stability criterion at two radial scale lengths is roughly  1000~\Msun pc$^{-2}$. This surface density corresponds to a disc {\it r}-band mass-to-light ratio $(M/L_d)_r= 2.88 (M/L)_{\odot}$. This mass-to-light ratio is  higher than the ratio obtained from the colour index of the disc and the model relations from \citet{Roediger2015}: $(M/L_d)_r=1.02$ and from \citet{Bell2003}: $(M/L_d)_r=1.7$.  However, a $(M/L_d)_r= 2.88$ is lower than an estimate of  $(M/L_d)_r=3.3$ based on a Salpeter stellar initial mass function, a stellar population age of 5~Gyr, a metallicity of [Fe/H]=$-0.3$~dex (see Fig.\ref{profiles_sp}) and allowing for a gas contribution. Thus, we conclude that the disc of UGC~1378  is close to marginal gravitational stability, suggesting that merger-induced strong gravitational perturbations did not occur after the formation of the HSB disc. 

\begin{figure*}

\includegraphics[width=0.9\textwidth,trim={0cm 3cm 0cm 3cm}]{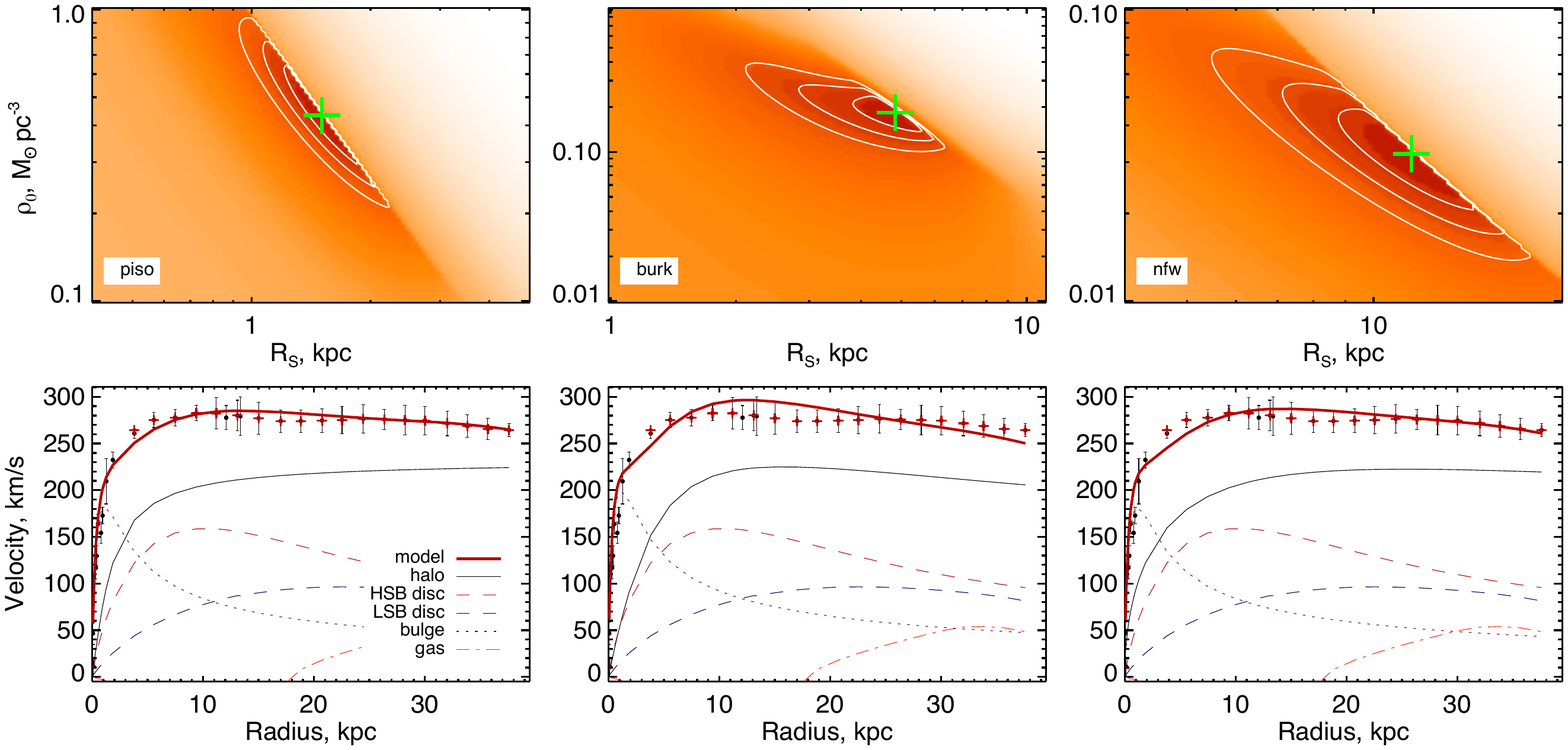}

\caption{Top panels: $\chi^2$ map for the parameters of dark halo, the darker the colour, the lower the $\chi^2$ and the better the fitting quality is.
The contours refer to $1\sigma$, $2\sigma$ and $3\sigma$ confidence limits.
The parameters corresponding to the $\chi^2$ minimum are shown by a cross in each map. Left column ~--- for the piso profile of the DM halo, centre ~--- to the Burkert profile, right ~--- for the NFW profile.
Bottom panels: the best-fitting models of the combined \HI (red crosses) + optical  (black dots) rotation curve.
}
\label{masmod}
\end{figure*}

During the modelling of the rotation curve, we fix the radial scales of the discs and the bulge from the {\it r}-band photometric parameters and the mass-to-light ratio of the HSB disc to 1.4, which follows from the   colour index  using the relations of \citet{Roediger2015}, \citet{Bell2003}. We allow the mass-to-light ratios of bulge and LSB disc to be free parameters. We summarize the results of the rotation curve mass-modelling in Fig.~\ref{masmod}. Following \citet{Saburova2016} we calculate $\chi^2$ maps for the dark halo parameters, also presented in Fig.~\ref{masmod}.  We give our fits to the structural components of UGC~1378 in Table~\ref{par}. The combined mass of the two discs is $2.9-3.8$ times less than the mass of the dark halo inside the LSB and \HI discs, higher than for normal HSB galaxies. However, inside the HSB disc, the ratio of dark to luminous matter is close to two as expected for normal galaxies \citep{Zasov2011}. The NFW halo scale radius is large (>10 kpc), but consistent with the large disc radius \citep[see, e.g.][]{Saburova2018}.

\begin{table*}
\begin{center}
\caption{The derived parameters of the main galaxy components with $1\sigma$ confidence limit errors. The columns contain the following data:
(1)~-- dark halo profile;
(2) and (3)~-- radial scale and central density of the DM halo;
(3)~-- mass of DM halo inside of radius 47 kpc;
(4)~-- disc {\it r} mass-to-light ratio;
(5)~-- central surface density of bulge
(6)~-- LSB and HSB discs mass\label{par};}
\renewcommand{\arraystretch}{1.5}
\begin{tabular}{lrlrl  rlrlrlrc}
\hline
dark halo	&	\multicolumn{2}{c}{$R_s$}&	\multicolumn{2}{c}{$\rho_0$ }&		\multicolumn{2}{c}{$M_{\rm halo}$}	 &	\multicolumn{2}{c}{$M/L$} &	\multicolumn{2}{c}{$(I_0)_b$}&\multicolumn{1}{c}{$M_{\rm discs}$}\\
&\multicolumn{2}{c}{kpc}&\multicolumn{2}{c}{$10^{-3}$ M$_{\odot}/$pc$^3$}&\multicolumn{2}{c}{$10^{10}$ M$_{\odot}$}&\multicolumn{2}{c}{M$_{\odot}/$L$_{\odot}$	}& \multicolumn{2}{c}{$10^{3}$ M$_{\odot}/$pc$^2$}&\multicolumn{1}{c}{$10^{10}$ M$_{\odot}$}\\
\hline
\hline
          Burkert&       4.83 & $^{+     0.36}_{-     0.45} $  &      183.38& $^{+     46.79}_{-      43.77} $  &       41.28& $^{+     4.27}_{-     14.85} $  &     1.08& $^{+      1.89}_{-      0.00} $  &     21.77& $^{+      0.00}_{-      0.00} $&14.4  \\                     NFW&    11.51& $^{+      2.94}_{-      2.04}$   &      31.99& $^{+      13.36}_{-      11.43}$     &         50.38& $^{+      5.34}_{-      16.14}$    &      1.08& $^{+      1.78}_{-      0.00}$    &     18.13&   $^{+      0.00}_{-      0.00}$&14.4  \\
                   piso&     1.51& $^{+     0.34}_{-     0.28} $  &      432.88& $^{+     209.93}_{-      134.51} $  &        55.18& $^{+     2.70}_{-     7.65} $  &     1.08& $^{+      0.82}_{-      0.00} $  &     18.13& $^{+      0.0}_{-      0.00} $&14.4  \\\hline
\hline
\end{tabular}
\end{center}
\end{table*}

\section{Discussion}\label{Discussion}

\begin{figure*}
\includegraphics[width=1.0\textwidth]{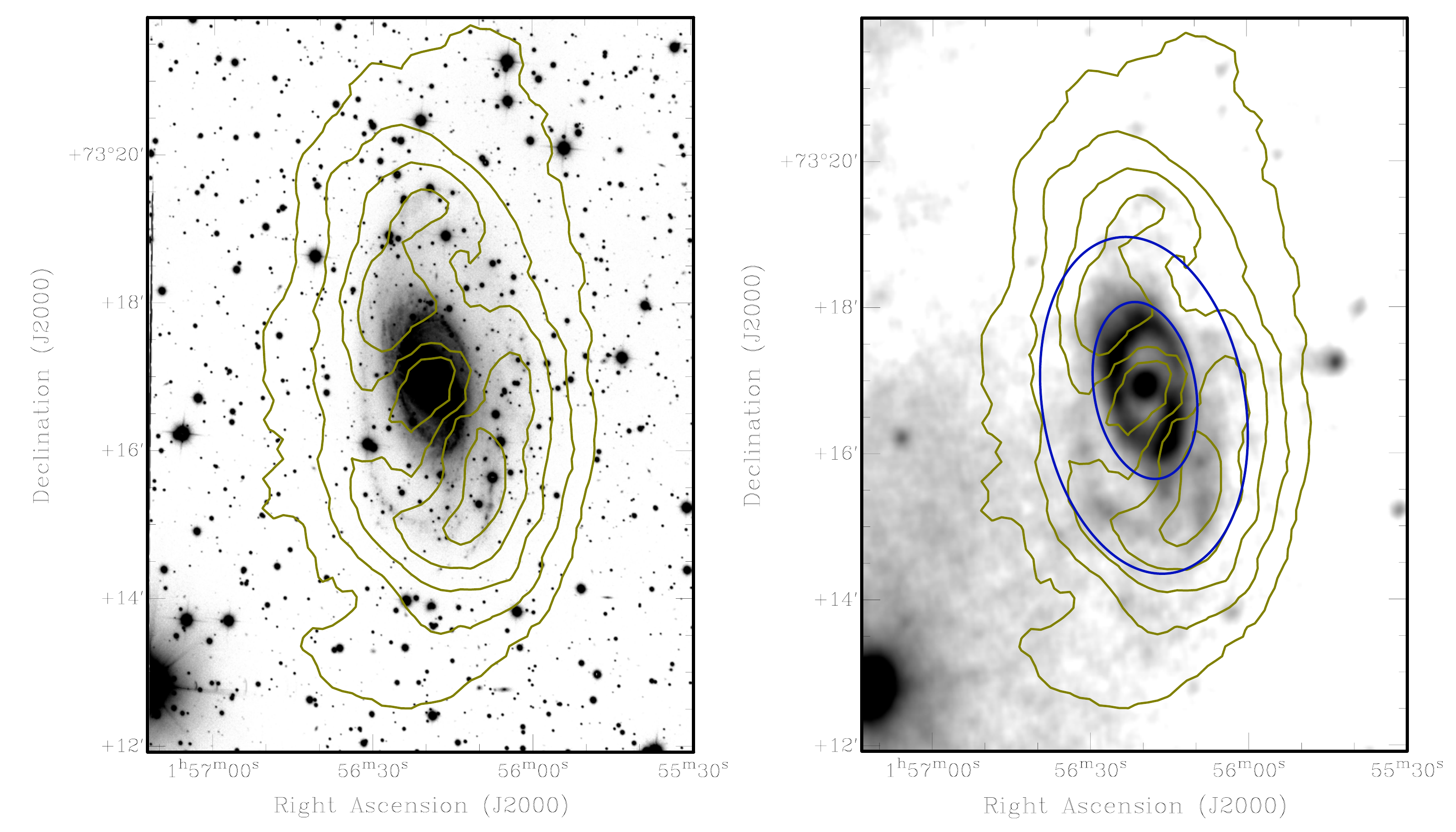}
\caption{The {\it r}-band  (left-hand column) and WISE band 3 (right-hand column) images  of UGC~1378 with overlaid contours of \HI density map taken from \citet{Mishra2017}  and areas used for the estimates of SFR of HSB and LSB discs  (central ellipse and a ring)}. \label{r_hi}
\end{figure*}

\begin{figure}
\vspace{-4.0cm}
\includegraphics[width=3\linewidth]{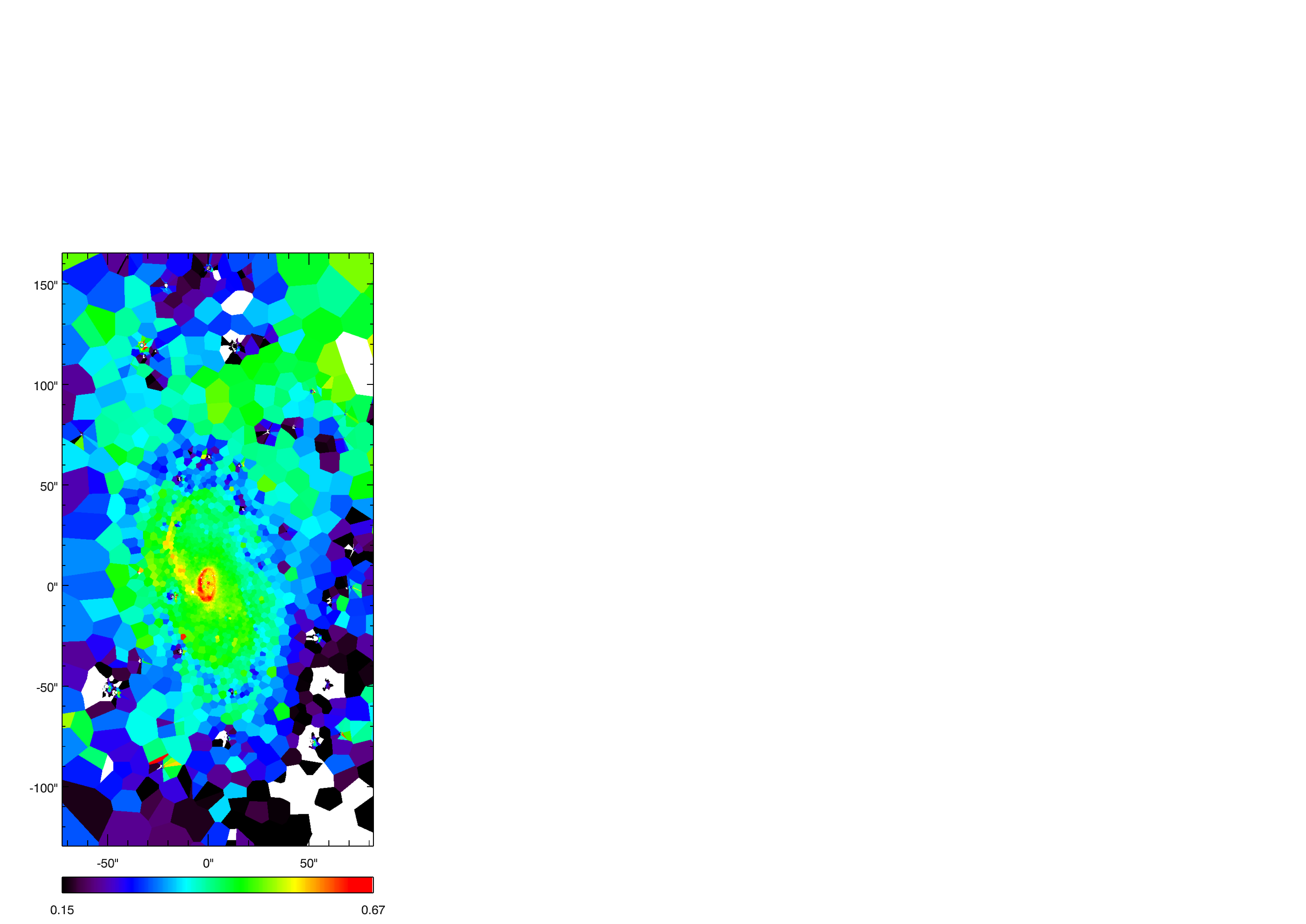}
\caption{The $g-r$ colour map corrected for Galactic extinction. We masked a region affected by a Galactic cirrus in the bottom of the map. North is up, east is left.}
\label{colour_map}
\end{figure}

UGC~1378 has a complex structure as evident in Fig.\ref{image} and from our photometric analysis, with a HSB inner disc and a LSB outer disc with prominent spiral arms. Fig.~\ref{r_hi}, where we demonstrate the WISE band 3 and {\it r}-band images of UGC~1378 with overlaid contours of \HI density map taken from \citet{Mishra2017}, shows that the HSB disc has roughly half the extent of the \HI disc, while the LSB disc extends to the edge of the \HI disc. The \HI contours trace the spiral arms visible in the {\it r}-band. The inner HSB region of UGC~1378 contains a large-scale bar with an embedded nuclear star-forming ring revealed by a circular red dust lane with a radius of 10~arcsec (2~kpc). The dust lane is visible in  Fig. \ref{colour_map} where we show the {\it g-r}  colour map of UGC~1378 obtained using adaptive binning to reach signal-to-noise level of 6  and corrected for Galactic extinction (assumed to be constant with radius). This ring is also visible in the stellar velocity dispersion profile (see Fig.~\ref{profiles_sp}) with two symmetric peaks that coincide with the edge of the ring. This structure is typical for barred galaxies and occurs in 20$\%$ of disc galaxies.  The ring may result from a recent gas inflow to the centre of the galaxy \citep{Comeron2010} but the ring is larger than in most cases described by \citet{Comeron2010}. The bar is surrounded by red spiral arms that form a ring-like structure. Our spectral analysis shows that the stellar population in the outer HSB region is metal-poor and is significantly younger than the bar and pseudo-bulge population.


\subsection{On the star formation rate of UGC~1378}

 The low surface brightness outer spiral arms have bluer $g-r$  colour (see Fig. \ref{colour_map}) which may indicate a young stellar population. Both the HSB disc and outer spirals are  clearly visible in WISE \citep{Wright2010} band~3 ($\lambda=12 \mu$m), see Fig.~\ref{r_hi}.  This image supports the presence of current star formation in the outer gLSB disc of UGC~1378.

We derive the star formation rate (SFR) for UGC~1378 by applying the relations from \citet{Jarrett2013} to the integrated UGC~1378 luminosity at 12 and 22~$\mu$m. The global SFR is estimated at between 1.2 and $2.3$~\Msun yr$^{-1}$.  This SFR is comparable or higher than the SFR in well-studied nearby galaxies \citep[see, e.g.][]{Jarrett2013} and is close to the Milky Way SFR \citep[see, f.e.,][]{Licquia2015, Chomiuk2011}.  However, accounting for the high \HI mass \citep{Mishra2017}, this SFR value is in good agreement with the \HI mass vs. SFR relation found by \citet{Boissier2008} for other LSB galaxies. They find LSB galaxies have a lower  SFR at a given gas mass than HSB galaxies. The SFR in UGC~1378 is also lower than expected from the SFR vs \HI mass relation for Local Volume galaxies \citep{Karachentsev2010}. However, the Local Volume galaxies do have much lower gas masses than UGC~1378.  

The mean SFR density is $(2\pm 0.1)\times 10^{-3}$ ~\Msun yr$^{-1}$ kpc$^{-2}$ which is higher than that observed in LSB galaxies \citep[see, e.g.][]{Wyder2009}. As expected, the LSB spiral arms in UGC~1378 have lower star formation rate than the HSB disc. The star forming ring (with the radius of approximately 50 arcsec) and bright centre have a SFR surface density comparable to the HSB disc: $(3.9\pm 0.3)\times 10^{-3}$ ~\Msun yr$^{-1}$ kpc$^{-2}$ and $(4.5\pm 0.4)\times 10^{-3}$ ~\Msun yr$^{-1}$ kpc$^{-2}$, respectively. We calculated the SFR density in the HSB disc: $(4\pm 0.3)\times 10^{-3}$ ~\Msun yr$^{-1}$ kpc$^{-2}$ and the  elliptical ring encompassing the LSB part: $(1\pm 0.1)\times 10^{-3}$ ~\Msun yr$^{-1}$ kpc$^{-2}$  (the regions used for SFR estimates of HSB and LSB discs are shown in Fig. \ref{r_hi}, right-hand column).  The SFR surface density in the UGC~1378 LSB spiral arms is higher than for most LSB galaxies studied by \citep{Wyder2009}.

\begin{figure}
\vspace{-28.5cm}
\hspace{-1.8cm}
\includegraphics[width=6.3\linewidth]{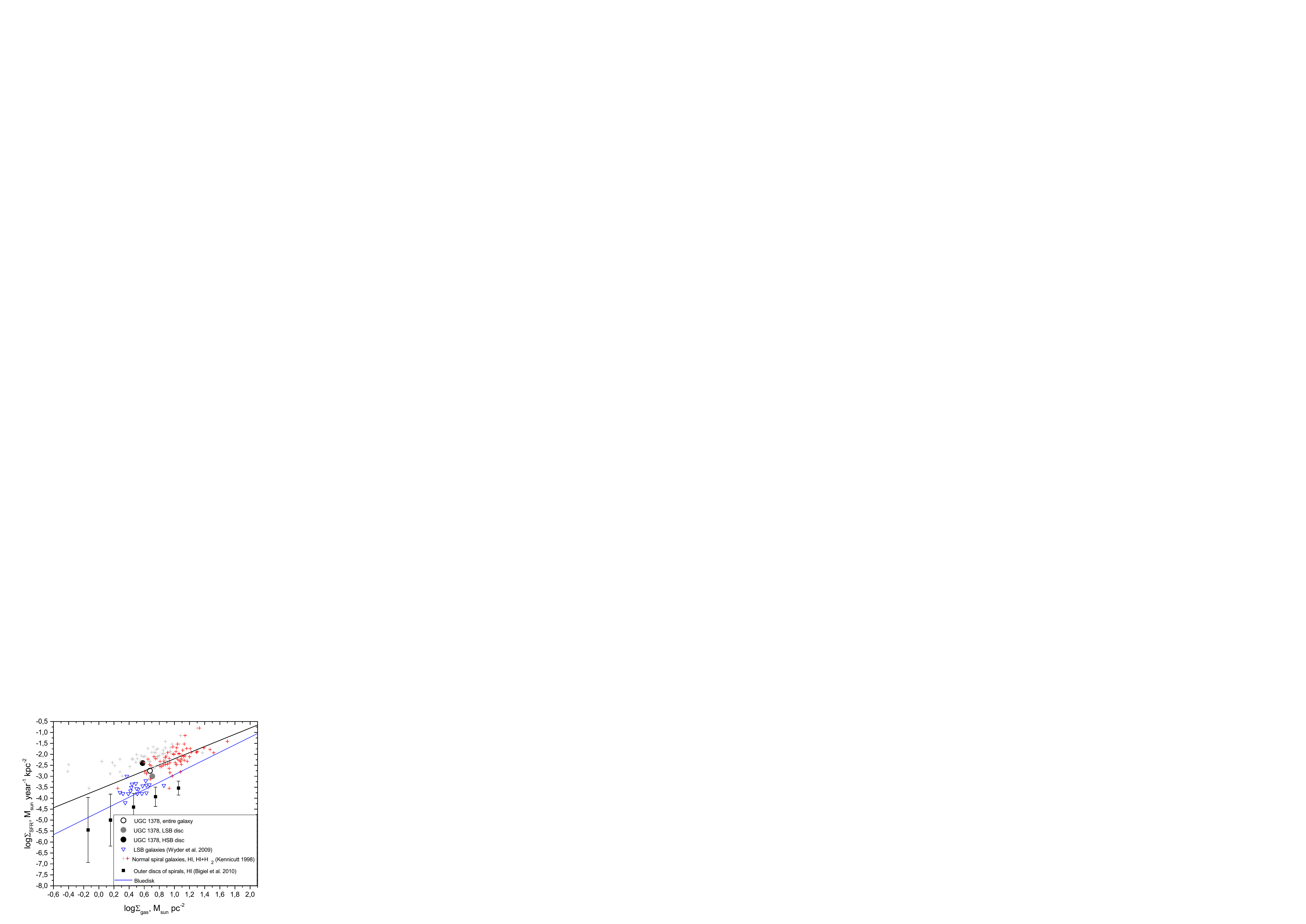}
\caption{The SFR surface density vs gas surface density. The LSB and HSB discs of UGC~1378 are shown as gray and black circles, respectively, and the  open circle denotes the mean value for the whole galaxy. The gas surface density for UGC~1378 is computed as the \HI surface density from the 0th moment map from \citet{Mishra2017} corrected for the helium abundance. The sample of normal spirals from \citet{Kennicutt1998} are plotted as   bright and faint crosses correspond to total and \HI gas densities, respectively. The black line corresponds to the fit by \citet{Kennicutt1998}.  The blue line shows the fit for the Bluedisk galaxies from \citet{Roychowdhury2015} for \HI surface density.  Triangles mark the position of LSB galaxies from the sample by \citet{Wyder2009} .  The estimates from \citet{Bigiel2010} for the outer parts of HSB spirals and \HI surface density are shown as squares.  }
\label{sfr}
\end{figure}

 In Fig.~\ref{sfr} we compare UGC~1378's SFR density vs gas surface density (the Schmidt--Kennicutt relation) to data in the literature. The gas surface density corresponds to \HI calculated from the 0th moment map  from \citet{Mishra2017}   in the same areas as SFR density.  Points for the HSB and LSB discs are plotted as black and gray circles, respectively. We plot the mean SFR and \HI surface density for the entire galaxy with a large open circle. The black line corresponds to the relation with an exponent of 1.4 found by \cite{Kennicutt1998}. Triangles give results for LSB galaxies published by \citep{Wyder2009}, and  bright and faint crosses show normal spiral galaxies from \citep{Kennicutt1998}  -- total and \HI surface densities.  A blue line shows the best-fitting relation for the \HI surface density of Bluedisk galaxies from \citet{Roychowdhury2015}. We also plot the SFR in the outer regions of spiral galaxies \citep{Bigiel2010} (square symbols).  In Fig.~\ref{sfr} the UGC~1378 measurements lie between normal spirals and LSB galaxies. The HSB disc data lie above the relation plotted for normal spirals, possibly indicating that the SFR is boosted by the bar driving gas to the star-forming rings. We cannot account for molecular gas since there are no available measurements for UGC~1378. The contribution of molecular gas would likely move the HSB disc of UGC~1378 toward the locus of normal galaxies.  Because the HSB SFR of UGC~1378 is close to the predicted SFR from the  \citet{Kennicutt1998} relation obtained from \HI densities (faint crosses in Fig. \ref{sfr}). The LSB disc of UGC~1378 lies below the correlation and accounting for molecular gas would only increase the deviation from the normal Schmidt--Kennicutt relation. Similar deviations are observed in ``classical'' LSB galaxies,   Bluedisk galaxies \citep{Roychowdhury2015}, outer parts of HSB spiral galaxies \citep{Bigiel2010} and \HI discs in early-type galaxies \citep{2017MNRAS.464..329Y}. These deviations for LSB galaxies are at least partially explained by their lower gas densities leading to lower SFRs \citep{Abramova2011}. A recent episode of gas accretion onto the disc of UGC~1378 may also contribute to a lower SFR if the gas is not yet fully participating in the star formation. \citet{Lutz2017} studied a sample of very \HI rich galaxies and proposed that very high specfic angular momentum in \HI rich galaxies prevents the accreted gas from being transported to the mid-plane of the disc and being converted into stars.  This mechanism may act to preserve giant gaseous discs. 

\subsection{On the formation scenario of UGC~1378}

The radius of the HSB disc of UGC~1378, $\sim$18 kpc, is comparable to the radius of the the Milky Way's disc. The normalized UGC~1378 SFR in this region is also comparable to the Milky Way SFR \citep[for a Milky Way star formation radius of 13.5 kpc][]{Chomiuk2011}. Hence, UGC~1378 resembles the Milky Way embedded in a gas-rich gLSB disc. Determining the formation scenario for this unusual system is an interesting challenge. The ``Bluedisk'' results indicate that H\,{\sc i}-rich galaxies in some repects similar to UGC~1378 may accrete gas from an extended gas reservoir \citep{Wang2015}. 

  \citet{Saburovaetal2018} suggested a three galaxy merger formation scenario for the giant LSB galaxy  UGC~1922, which also have giant gaseous discs. The {\sc galmer} simulations \citep{Chilingarian+10} of a major in-plane merger of two gas-rich bulgeless (e.g. Sd) galaxies also sometimes result in formation of a giant thin disc which might be classified as a gLSB. However, because the HSB disc of UGC~1378 is close to marginal gravitational stability, such a catastrophic formation scenario for UGC~1378 is unlikely, because otherwise the major merger would have heated the disc and increased stellar velocity dispersion.  The unusually large radius of UGC~1378's gaseous disc more likely results from two factors: a rich gas reservoir and an isolated environment that protects the disc from destruction by merger events.
   The fact that we observe UGC~1378 in isolation can indicate that it has been living in low density environment from the time of its formation and primary mass assembly because no mechanism can eject a large dark halo hosting a massive galaxy from a cluster or a rich group to a low density region of the Universe. 
  
  The blue colour and clumpy structure of the LSB spirals (see Fig.~\ref{colour_map}) suggest a recent burst of star formation in the LSB disc.  The origins of a star burst in this low density region is puzzling, but may be the result of a recent minor merger.  The variable obscuration in the foreground of UGC~1378 may hide traces of a putative merger event in our deep imaging data. Alternatively, the blue colour of the LSB disc may result from its low metallicity, consistent with slow accretion of gas from a cosmic filament.
 
 The formation histories of well studied gLSB galaxies appear to be diverse if we include  galaxies  with properties intermediate between gLSB and HSB, particularly those with slightly smaller disc sizes like UGC~1378.  While major mergers  could be the likely origin of the most extended gLSBs like Malin~1 or UGC~1922 \citep{Zhu2018, Saburovaetal2018}, this is not the case for UGC~1378. 
 
 \section{Conclusions}\label{conclusion} 

We describe deep long-slit spectroscopy and imaging for UGC~1378, a galaxy with a giant \HI and LSB disc. UGC~1378's structure is complex, including a HSB pseudo-bulge and an HSB disc with a bar, rings and spiral arms, and a gas-rich gLSB disc with spiral arms and a radius of 50~kpc. We model UGC~1378's stellar population and rotation curve and conclude that:
\begin{enumerate}
 
\item  UGC~1378 resides between normal spirals and LSB galaxies in a SFR surface density vs gas surface density diagram. The LSB disc has lower SFR surface density than expected from its gas density  as for most gLSBs \citep{Wyder2009} which may indicate recent gas accretion. The global SFR for UGC~1378 ($1.2 \dots 2.3$~\Msun yr$^{-1}$) is close to that of the Milky Way.
\item The stellar velocity dispersion of the HSB disc is close to that expected for its marginal gravitational stability, inconsistent with recent major merger events.
\item The dark matter halo dominates the mass inside the giant LSB disc radius.  Within the HSB part of the galaxy the ratio or dark to luminous  mass is close to 2 as expected for normal spirals. 
\item  We suggest that the formation of UGC~1378 occurred in two stages. The first stage, in common with MW-type galaxies,
included several episodes of merging during which the HSB part formed and likely most satellites were accreted. In the second stage gLSB stellar and gaseous discs were formed by accretion of metal-poor gas from a cosmic filament. The unusually large radius of UGC~1378's gaseous disc could result from the presence of a gas reservoir (e.g. a cosmic filament). A low density environment likely helped to preserve the giant disc.
\end{enumerate}

\section*{Acknowledgements}  
We are grateful to the anonymous referee for helpful comments.
AS acknowledges  The Russian Science Foundation (RSCF) grant  19-72-20089  that supported research on dynamical modelling of gLSBs and Russian Foundation for Basic Research grant 18-32-20120 that supported observational studies of star formation in disc galaxies.  IC,  AK and IK acknowledge the RScF grant 17-72-20119 for the reduction and analysis of spectral and photometric data. AS, IC, IK, and AK acknowledge the Leading Scientific School in astrophysics (direction: extragalactic astronomy) at Moscow State University. The authors are grateful to the staff of the MMT Observatory and of the Telescope Data Center at Smithsonian Astrophysical Observatory.

\bibliographystyle{mnras}
\bibliography{u1378}

\begin{thebibliography}{}
\makeatletter
\relax
\def\mn@urlcharsother{\let\do\@makeother \do\$\do\&\do\#\do\^\do\_\do\%\do\~}
\def\mn@doi{\begingroup\mn@urlcharsother \@ifnextchar [ {\mn@doi@}
  {\mn@doi@[]}}
\def\mn@doi@[#1]#2{\def\@tempa{#1}\ifx\@tempa\@empty \href
  {http://dx.doi.org/#2} {doi:#2}\else \href {http://dx.doi.org/#2} {#1}\fi
  \endgroup}
\def\mn@eprint#1#2{\mn@eprint@#1:#2::\@nil}
\def\mn@eprint@arXiv#1{\href {http://arxiv.org/abs/#1} {{\tt arXiv:#1}}}
\def\mn@eprint@dblp#1{\href {http://dblp.uni-trier.de/rec/bibtex/#1.xml}
  {dblp:#1}}
\def\mn@eprint@#1:#2:#3:#4\@nil{\def\@tempa {#1}\def\@tempb {#2}\def\@tempc
  {#3}\ifx \@tempc \@empty \let \@tempc \@tempb \let \@tempb \@tempa \fi \ifx
  \@tempb \@empty \def\@tempb {arXiv}\fi \@ifundefined
  {mn@eprint@\@tempb}{\@tempb:\@tempc}{\expandafter \expandafter \csname
  mn@eprint@\@tempb\endcsname \expandafter{\@tempc}}}

\bibitem[\protect\citeauthoryear{{Abramova} \& {Zasov}}{{Abramova} \&
  {Zasov}}{2011}]{Abramova2011}
{Abramova} O.~V.,  {Zasov} A.~V.,  2011, \mn@doi [Astronomy Reports]
  {10.1134/S1063772911030012}, \href
  {http://adsabs.harvard.edu/abs/2011ARep...55..202A} {55, 202}

\bibitem[\protect\citeauthoryear{{Afanasiev} \& {Moiseev}}{{Afanasiev} \&
  {Moiseev}}{2005}]{AfanasievMoiseev2005}
{Afanasiev} V.~L.,  {Moiseev} A.~V.,  2005, \mn@doi [Astronomy Letters]
  {10.1134/1.1883351}, \href
  {http://adsabs.harvard.edu/abs/2005AstL...31..194A} {31, 194}

\bibitem[\protect\citeauthoryear{{Bell}, {McIntosh}, {Katz}  \&
  {Weinberg}}{{Bell} et~al.}{2003}]{Bell2003}
{Bell} E.~F.,  {McIntosh} D.~H.,  {Katz} N.,   {Weinberg} M.~D.,  2003, \mn@doi
  [\apjs] {10.1086/378847}, \href
  {http://adsabs.harvard.edu/abs/2003ApJS..149..289B} {149, 289}

\bibitem[\protect\citeauthoryear{{Bertin}, {Mellier}, {Radovich}, {Missonnier},
  {Didelon}  \& {Morin}}{{Bertin} et~al.}{2002}]{swarp}
{Bertin} E.,  {Mellier} Y.,  {Radovich} M.,  {Missonnier} G.,  {Didelon} P.,
  {Morin} B.,  2002, in {Bohlender} D.~A.,  {Durand} D.,   {Handley} T.~H.,
  eds,  Astronomical Society of the Pacific Conference Series Vol. 281,
  Astronomical Data Analysis Software and Systems XI. p.~228

\bibitem[\protect\citeauthoryear{{Bigiel}, {Leroy}, {Walter}, {Blitz},
  {Brinks}, {de Blok}  \& {Madore}}{{Bigiel} et~al.}{2010}]{Bigiel2010}
{Bigiel} F.,  {Leroy} A.,  {Walter} F.,  {Blitz} L.,  {Brinks} E.,  {de Blok}
  W.~J.~G.,   {Madore} B.,  2010, \mn@doi [\aj] {10.1088/0004-6256/140/5/1194},
  \href {https://ui.adsabs.harvard.edu/abs/2010AJ....140.1194B} {140, 1194}

\bibitem[\protect\citeauthoryear{{Boissier} et~al.,}{{Boissier}
  et~al.}{2008}]{Boissier2008}
{Boissier} S.,  et~al., 2008, \mn@doi [\apj] {10.1086/588580}, \href
  {https://ui.adsabs.harvard.edu/abs/2008ApJ...681..244B} {681, 244}

\bibitem[\protect\citeauthoryear{{Boissier} et~al.,}{{Boissier}
  et~al.}{2016}]{Boissier2016}
{Boissier} S.,  et~al., 2016, \mn@doi [\aap] {10.1051/0004-6361/201629226},
  \href {http://adsabs.harvard.edu/abs/2016A%26A...593A.126B} {593, A126}

\bibitem[\protect\citeauthoryear{{Burkert}}{{Burkert}}{1995}]{Burkert1995}
{Burkert} A.,  1995, \mn@doi [\apjl] {10.1086/309560}, \href
  {http://adsabs.harvard.edu/abs/1995ApJ...447L..25B} {447, L25}

\bibitem[\protect\citeauthoryear{{Cappellari} \& {Emsellem}}{{Cappellari} \&
  {Emsellem}}{2004}]{CE04}
{Cappellari} M.,  {Emsellem} E.,  2004, \mn@doi [\pasp] {10.1086/381875}, 116,
  138

\bibitem[\protect\citeauthoryear{{Chambers} et~al.,}{{Chambers}
  et~al.}{2016}]{Chambers2016_panstarrs1}
{Chambers} K.~C.,  et~al., 2016, preprint, \href
  {https://ui.adsabs.harvard.edu/#abs/2016arXiv161205560C} {p.
  arXiv:1612.05560} (\mn@eprint {arXiv} {1612.05560})

\bibitem[\protect\citeauthoryear{{Chilingarian}, {Prugniel}, {Sil'Chenko}  \&
  {Koleva}}{{Chilingarian} et~al.}{2007a}]{Chilingarian2007a}
{Chilingarian} I.,  {Prugniel} P.,  {Sil'Chenko} O.,   {Koleva} M.,  2007a, in
  {Vazdekis} A.,  {Peletier} R.,  eds,  IAU Symposium Vol. 241, IAU Symposium.
  pp 175--176 (\mn@eprint {arXiv} {0709.3047}),
  \mn@doi{10.1017/S1743921307007752}

\bibitem[\protect\citeauthoryear{{Chilingarian}, {Prugniel}, {Sil'Chenko}  \&
  {Afanasiev}}{{Chilingarian} et~al.}{2007b}]{Chilingarian2007b}
{Chilingarian} I.~V.,  {Prugniel} P.,  {Sil'Chenko} O.~K.,   {Afanasiev} V.~L.,
   2007b, \mn@doi [\mnras] {10.1111/j.1365-2966.2007.11549.x}, \href
  {http://adsabs.harvard.edu/abs/2007MNRAS.376.1033C} {376, 1033}

\bibitem[\protect\citeauthoryear{{Chilingarian}, {Novikova}, {Cayatte},
  {Combes}, {Di Matteo}  \& {Zasov}}{{Chilingarian}
  et~al.}{2009}]{Chilingarian2009}
{Chilingarian} I.~V.,  {Novikova} A.~P.,  {Cayatte} V.,  {Combes} F.,  {Di
  Matteo} P.,   {Zasov} A.~V.,  2009, \mn@doi [\aap]
  {10.1051/0004-6361/200911684}, \href
  {http://adsabs.harvard.edu/abs/2009A%26A...504..389C} {504, 389}

\bibitem[\protect\citeauthoryear{{Chilingarian}, {Di Matteo}, {Combes},
  {Melchior}  \& {Semelin}}{{Chilingarian} et~al.}{2010}]{Chilingarian+10}
{Chilingarian} I.~V.,  {Di Matteo} P.,  {Combes} F.,  {Melchior} A.-L.,
  {Semelin} B.,  2010, \mn@doi [\aap] {10.1051/0004-6361/200912938}, 518, A61

\bibitem[\protect\citeauthoryear{{Chomiuk} \& {Povich}}{{Chomiuk} \&
  {Povich}}{2011}]{Chomiuk2011}
{Chomiuk} L.,  {Povich} M.~S.,  2011, \mn@doi [\aj]
  {10.1088/0004-6256/142/6/197}, \href
  {http://adsabs.harvard.edu/abs/2011AJ....142..197C} {142, 197}

\bibitem[\protect\citeauthoryear{{Chung} \& {Bureau}}{{Chung} \&
  {Bureau}}{2004}]{Chung2004}
{Chung} A.,  {Bureau} M.,  2004, \mn@doi [\aj] {10.1086/420988}, \href
  {http://adsabs.harvard.edu/abs/2004AJ....127.3192C} {127, 3192}

\bibitem[\protect\citeauthoryear{{Comer{\'o}n}, {Knapen}, {Beckman},
  {Laurikainen}, {Salo}, {Mart{\'{\i}}nez-Valpuesta}  \& {Buta}}{{Comer{\'o}n}
  et~al.}{2010}]{Comeron2010}
{Comer{\'o}n} S.,  {Knapen} J.~H.,  {Beckman} J.~E.,  {Laurikainen} E.,  {Salo}
  H.,  {Mart{\'{\i}}nez-Valpuesta} I.,   {Buta} R.~J.,  2010, \mn@doi [\mnras]
  {10.1111/j.1365-2966.2009.16057.x}, \href
  {http://adsabs.harvard.edu/abs/2010MNRAS.402.2462C} {402, 2462}

\bibitem[\protect\citeauthoryear{{Deeley} et~al.,}{{Deeley}
  et~al.}{2017}]{Deeley2017}
{Deeley} S.,  et~al., 2017, \mn@doi [\mnras] {10.1093/mnras/stx441}, \href
  {http://adsabs.harvard.edu/abs/2017MNRAS.467.3934D} {467, 3934}

\bibitem[\protect\citeauthoryear{{Fabricant} et~al.,}{{Fabricant}
  et~al.}{2019}]{Fabricant2019}
{Fabricant} D.,  et~al., 2019, arXiv e-prints, \href
  {http://adsabs.harvard.edu/abs/2019arXiv190503320F} {}

\bibitem[\protect\citeauthoryear{{Fraternali}, {Oosterloo}, {Sancisi}  \& {van
  Moorsel}}{{Fraternali} et~al.}{2001}]{Fraternali2001}
{Fraternali} F.,  {Oosterloo} T.,  {Sancisi} R.,   {van Moorsel} G.,  2001,
  \mn@doi [\apjl] {10.1086/338102}, \href
  {http://adsabs.harvard.edu/abs/2001ApJ...562L..47F} {562, L47}

\bibitem[\protect\citeauthoryear{{Fraternali}, {van Moorsel}, {Sancisi}  \&
  {Oosterloo}}{{Fraternali} et~al.}{2002}]{Fraternali2002}
{Fraternali} F.,  {van Moorsel} G.,  {Sancisi} R.,   {Oosterloo} T.,  2002,
  \mn@doi [\aj] {10.1086/340358}, \href
  {http://adsabs.harvard.edu/abs/2002AJ....123.3124F} {123, 3124}

\bibitem[\protect\citeauthoryear{{Freeman}}{{Freeman}}{1970}]{Freeman1970}
{Freeman} K.~C.,  1970, \mn@doi [\apj] {10.1086/150474}, \href
  {https://ui.adsabs.harvard.edu/abs/1970ApJ...160..811F} {160, 811}

\bibitem[\protect\citeauthoryear{{Galaz}, {Milovic}, {Suc}, {Busta}, {Lizana},
  {Infante}  \& {Royo}}{{Galaz} et~al.}{2015}]{Galaz2015}
{Galaz} G.,  {Milovic} C.,  {Suc} V.,  {Busta} L.,  {Lizana} G.,  {Infante} L.,
    {Royo} S.,  2015, \mn@doi [\apjl] {10.1088/2041-8205/815/2/L29}, \href
  {https://ui.adsabs.harvard.edu/abs/2015ApJ...815L..29G} {815, L29}

\bibitem[\protect\citeauthoryear{{Ger{\'e}b}, {Janowiecki}, {Catinella},
  {Cortese}  \& {Kilborn}}{{Ger{\'e}b} et~al.}{2018}]{Gereb2018}
{Ger{\'e}b} K.,  {Janowiecki} S.,  {Catinella} B.,  {Cortese} L.,   {Kilborn}
  V.,  2018, \mn@doi [\mnras] {10.1093/mnras/sty214}, \href
  {http://adsabs.harvard.edu/abs/2018MNRAS.476..896G} {476, 896}

\bibitem[\protect\citeauthoryear{{Hagen} et~al.,}{{Hagen}
  et~al.}{2016}]{Hagen2016}
{Hagen} L. M.~Z.,  et~al., 2016, \mn@doi [\apj] {10.3847/0004-637X/826/2/210},
  \href {https://ui.adsabs.harvard.edu/abs/2016ApJ...826..210H} {826, 210}

\bibitem[\protect\citeauthoryear{{Huang} et~al.,}{{Huang}
  et~al.}{2014}]{highmass}
{Huang} S.,  et~al., 2014, \mn@doi [\apj] {10.1088/0004-637X/793/1/40}, \href
  {http://adsabs.harvard.edu/abs/2014ApJ...793...40H} {793, 40}

\bibitem[\protect\citeauthoryear{{Jarrett} et~al.,}{{Jarrett}
  et~al.}{2013}]{Jarrett2013}
{Jarrett} T.~H.,  et~al., 2013, \mn@doi [\aj] {10.1088/0004-6256/145/1/6},
  \href {http://adsabs.harvard.edu/abs/2013AJ....145....6J} {145, 6}

\bibitem[\protect\citeauthoryear{{Jedrzejewski}}{{Jedrzejewski}}{1987}]{Jedrzejewski1987}
{Jedrzejewski} R.~I.,  1987, \mn@doi [\mnras] {10.1093/mnras/226.4.747}, \href
  {http://adsabs.harvard.edu/abs/1987MNRAS.226..747J} {226, 747}

\bibitem[\protect\citeauthoryear{{Jester} et~al.,}{{Jester}
  et~al.}{2005}]{Jester2005}
{Jester} S.,  et~al., 2005, \mn@doi [\aj] {10.1086/432466}, \href
  {https://ui.adsabs.harvard.edu/abs/2005AJ....130..873J} {130, 873}

\bibitem[\protect\citeauthoryear{{J{\'o}zsa}}{{J{\'o}zsa}}{2007}]{J2007}
{J{\'o}zsa} G.~I.~G.,  2007, \mn@doi [\aap] {10.1051/0004-6361:20066165}, \href
  {http://adsabs.harvard.edu/abs/2007A%26A...468..903J} {468, 903}

\bibitem[\protect\citeauthoryear{{Kansky} et~al.,}{{Kansky}
  et~al.}{2019}]{Kansky2019}
{Kansky} J.,  et~al., 2019, arXiv e-prints, \href
  {http://adsabs.harvard.edu/abs/2019arXiv190503321K} {}

\bibitem[\protect\citeauthoryear{{Karachentsev} \& {Kaisin}}{{Karachentsev} \&
  {Kaisin}}{2010}]{Karachentsev2010}
{Karachentsev} I.~D.,  {Kaisin} S.~S.,  2010, \mn@doi [\aj]
  {10.1088/0004-6256/140/5/1241}, \href
  {https://ui.adsabs.harvard.edu/abs/2010AJ....140.1241K} {140, 1241}

\bibitem[\protect\citeauthoryear{{Kasparova}, {Saburova}, {Katkov},
  {Chilingarian}  \& {Bizyaev}}{{Kasparova} et~al.}{2014}]{Kasparova2014}
{Kasparova} A.~V.,  {Saburova} A.~S.,  {Katkov} I.~Y.,  {Chilingarian} I.~V.,
  {Bizyaev} D.~V.,  2014, \mn@doi [\mnras] {10.1093/mnras/stt1982}, \href
  {http://adsabs.harvard.edu/abs/2014MNRAS.437.3072K} {437, 3072}

\bibitem[\protect\citeauthoryear{{Katkov} \& {Chilingarian}}{{Katkov} \&
  {Chilingarian}}{2011}]{KatkovChilingarian2011}
{Katkov} I.~Y.,  {Chilingarian} I.~V.,  2011, in {Evans} I.~N.,  {Accomazzi}
  A.,  {Mink} D.~J.,   {Rots} A.~H.,  eds,  Astronomical Society of the Pacific
  Conference Series Vol. 442, Astronomical Data Analysis Software and Systems
  XX. p.~143 (\mn@eprint {arXiv} {1012.4125})

\bibitem[\protect\citeauthoryear{{Kennicutt}}{{Kennicutt}}{1998}]{Kennicutt1998}
{Kennicutt} Jr. R.~C.,  1998, \mn@doi [\apj] {10.1086/305588}, \href
  {http://adsabs.harvard.edu/abs/1998ApJ...498..541K} {498, 541}

\bibitem[\protect\citeauthoryear{{Khoperskov}, {Zasov}  \&
  {Tyurina}}{{Khoperskov} et~al.}{2003}]{Khoperskov2003}
{Khoperskov} A.~V.,  {Zasov} A.~V.,   {Tyurina} N.~V.,  2003, \mn@doi
  [Astronomy Reports] {10.1134/1.1575851}, \href
  {http://adsabs.harvard.edu/abs/2003ARep...47..357K} {47, 357}

\bibitem[\protect\citeauthoryear{{Kirby}, {Rizzi}, {Held}, {Cohen}, {Cole},
  {Manning}, {Skillman}  \& {Weisz}}{{Kirby} et~al.}{2016}]{Kirbyetal2016}
{Kirby} E.~N.,  {Rizzi} L.,  {Held} E.~V.,  {Cohen} J.~G.,  {Cole} A.~A.,
  {Manning} E.~M.,  {Skillman} E.~D.,   {Weisz} D.~R.,  2016, preprint, \href
  {http://adsabs.harvard.edu/abs/2016arXiv161008505K} {} (\mn@eprint {arXiv}
  {1610.08505})

\bibitem[\protect\citeauthoryear{{Kormendy} \& {Kennicutt}}{{Kormendy} \&
  {Kennicutt}}{2004}]{Kormendy2004}
{Kormendy} J.,  {Kennicutt} Jr. R.~C.,  2004, \mn@doi [\araa]
  {10.1146/annurev.astro.42.053102.134024}, \href
  {http://adsabs.harvard.edu/abs/2004ARA%26A..42..603K} {42, 603}

\bibitem[\protect\citeauthoryear{{Kregel} \& {van der Kruit}}{{Kregel} \& {van
  der Kruit}}{2004}]{Kregel2004}
{Kregel} M.,  {van der Kruit} P.~C.,  2004, \mn@doi [\mnras]
  {10.1111/j.1365-2966.2004.08307.x}, \href
  {https://ui.adsabs.harvard.edu/abs/2004MNRAS.355..143K} {355, 143}

\bibitem[\protect\citeauthoryear{{Larson}, {Tinsley}  \& {Caldwell}}{{Larson}
  et~al.}{1980}]{Larson1980}
{Larson} R.~B.,  {Tinsley} B.~M.,   {Caldwell} C.~N.,  1980, \mn@doi [\apj]
  {10.1086/157917}, \href {http://adsabs.harvard.edu/abs/1980ApJ...237..692L}
  {237, 692}

\bibitem[\protect\citeauthoryear{{Le Borgne}, {Rocca-Volmerange}, {Prugniel},
  {Lan{\c c}on}, {Fioc}  \& {Soubiran}}{{Le Borgne}
  et~al.}{2004}]{LeBorgneetal2004}
{Le Borgne} D.,  {Rocca-Volmerange} B.,  {Prugniel} P.,  {Lan{\c c}on} A.,
  {Fioc} M.,   {Soubiran} C.,  2004, \mn@doi [\aap]
  {10.1051/0004-6361:200400044}, \href
  {http://adsabs.harvard.edu/abs/2004A%26A...425..881L} {425, 881}

\bibitem[\protect\citeauthoryear{{Licquia} \& {Newman}}{{Licquia} \&
  {Newman}}{2015}]{Licquia2015}
{Licquia} T.~C.,  {Newman} J.~A.,  2015, \mn@doi [\apj]
  {10.1088/0004-637X/806/1/96}, \href
  {http://adsabs.harvard.edu/abs/2015ApJ...806...96L} {806, 96}

\bibitem[\protect\citeauthoryear{{Lutz} et~al.,}{{Lutz}
  et~al.}{2017}]{Lutz2017}
{Lutz} K.~A.,  et~al., 2017, \mn@doi [\mnras] {10.1093/mnras/stx053}, \href
  {http://adsabs.harvard.edu/abs/2017MNRAS.467.1083L} {467, 1083}

\bibitem[\protect\citeauthoryear{{Mishra}, {Kantharia}, {Das}, {Omar}  \&
  {Srivastava}}{{Mishra} et~al.}{2017}]{Mishra2017}
{Mishra} A.,  {Kantharia} N.~G.,  {Das} M.,  {Omar} A.,   {Srivastava} D.~C.,
  2017, \mn@doi [\mnras] {10.1093/mnras/stw2506}, \href
  {http://adsabs.harvard.edu/abs/2017MNRAS.464.2741M} {464, 2741}

\bibitem[\protect\citeauthoryear{{Navarro}, {Frenk}  \& {White}}{{Navarro}
  et~al.}{1996}]{nfw1996}
{Navarro} J.~F.,  {Frenk} C.~S.,   {White} S.~D.~M.,  1996, \mn@doi [\apj]
  {10.1086/177173}, \href {http://adsabs.harvard.edu/abs/1996ApJ462563N} {462,
  563}

\bibitem[\protect\citeauthoryear{{O'Neil}, {Bothun}  \& {Cornell}}{{O'Neil}
  et~al.}{1997}]{obc1997}
{O'Neil} K.,  {Bothun} G.~D.,   {Cornell} M.~E.,  1997, \mn@doi [\aj]
  {10.1086/118338}, \href
  {https://ui.adsabs.harvard.edu/abs/1997AJ....113.1212O} {113, 1212}

\bibitem[\protect\citeauthoryear{{Oosterloo}, {Fraternali}  \&
  {Sancisi}}{{Oosterloo} et~al.}{2007}]{Oosterloo2007}
{Oosterloo} T.,  {Fraternali} F.,   {Sancisi} R.,  2007, \mn@doi [\aj]
  {10.1086/520332}, \href {http://adsabs.harvard.edu/abs/2007AJ....134.1019O}
  {134, 1019}

\bibitem[\protect\citeauthoryear{{Pe{\~n}arrubia}, {McConnachie}  \&
  {Babul}}{{Pe{\~n}arrubia} et~al.}{2006}]{2006ApJ650L33P}
{Pe{\~n}arrubia} J.,  {McConnachie} A.,   {Babul} A.,  2006, \mn@doi [\apjl]
  {10.1086/508656}, \href {http://adsabs.harvard.edu/abs/2006ApJ...650L..33P}
  {650, L33}

\bibitem[\protect\citeauthoryear{{Peng}, {Ho}, {Impey}  \& {Rix}}{{Peng}
  et~al.}{2002}]{Peng2002}
{Peng} C.~Y.,  {Ho} L.~C.,  {Impey} C.~D.,   {Rix} H.-W.,  2002, \mn@doi [\aj]
  {10.1086/340952}, \href {http://adsabs.harvard.edu/abs/2002AJ....124..266P}
  {124, 266}

\bibitem[\protect\citeauthoryear{{Rodriguez-Gomez} et~al.,}{{Rodriguez-Gomez}
  et~al.}{2015}]{Rodriguez-Gomez2015}
{Rodriguez-Gomez} V.,  et~al., 2015, \mn@doi [\mnras] {10.1093/mnras/stv264},
  \href {http://adsabs.harvard.edu/abs/2015MNRAS.449...49R} {449, 49}

\bibitem[\protect\citeauthoryear{{Roediger} \& {Courteau}}{{Roediger} \&
  {Courteau}}{2015}]{Roediger2015}
{Roediger} J.~C.,  {Courteau} S.,  2015, \mn@doi [\mnras]
  {10.1093/mnras/stv1499}, \href
  {http://adsabs.harvard.edu/abs/2015MNRAS.452.3209R} {452, 3209}

\bibitem[\protect\citeauthoryear{{Romeo} \& {Wiegert}}{{Romeo} \&
  {Wiegert}}{2011}]{Romeo2011}
{Romeo} A.~B.,  {Wiegert} J.,  2011, \mn@doi [\mnras]
  {10.1111/j.1365-2966.2011.19120.x}, \href
  {http://adsabs.harvard.edu/abs/2011MNRAS.416.1191R} {416, 1191}

\bibitem[\protect\citeauthoryear{{Roychowdhury}, {Huang}, {Kauffmann}, {Wang}
  \& {Chengalur}}{{Roychowdhury} et~al.}{2015}]{Roychowdhury2015}
{Roychowdhury} S.,  {Huang} M.-L.,  {Kauffmann} G.,  {Wang} J.,   {Chengalur}
  J.~N.,  2015, \mn@doi [\mnras] {10.1093/mnras/stv515}, \href
  {https://ui.adsabs.harvard.edu/abs/2015MNRAS.449.3700R} {449, 3700}

\bibitem[\protect\citeauthoryear{{Saburova}}{{Saburova}}{2011}]{Saburova2011}
{Saburova} A.~S.,  2011, \mn@doi [Astronomy Reports]
  {10.1134/S1063772911050064}, \href
  {http://adsabs.harvard.edu/abs/2011ARep...55..409S} {55, 409}

\bibitem[\protect\citeauthoryear{{Saburova}}{{Saburova}}{2018}]{Saburova2018}
{Saburova} A.~S.,  2018, \mn@doi [\mnras] {10.1093/mnras/stx2583}, \href
  {http://adsabs.harvard.edu/abs/2018MNRAS.473.3796S} {473, 3796}

\bibitem[\protect\citeauthoryear{{Saburova} \& {Zasov}}{{Saburova} \&
  {Zasov}}{2013}]{Saburova2013}
{Saburova} A.~S.,  {Zasov} A.~V.,  2013, \mn@doi [Astronomische Nachrichten]
  {10.1002/asna.201311922}, \href
  {http://adsabs.harvard.edu/abs/2013AN....334..785S} {334, 785}

\bibitem[\protect\citeauthoryear{{Saburova}, {Kasparova}  \&
  {Katkov}}{{Saburova} et~al.}{2016}]{Saburova2016}
{Saburova} A.~S.,  {Kasparova} A.~V.,   {Katkov} I.~Y.,  2016, \mn@doi [\mnras]
  {10.1093/mnras/stw2040}, \href
  {http://adsabs.harvard.edu/abs/2016MNRAS.463.2523S} {463, 2523}

\bibitem[\protect\citeauthoryear{{Saburova}, {Katkov}, {Khoperskov}, {Zasov}
  \& {Uklein}}{{Saburova} et~al.}{2017}]{Saburova2017}
{Saburova} A.~S.,  {Katkov} I.~Y.,  {Khoperskov} S.~A.,  {Zasov} A.~V.,
  {Uklein} R.~I.,  2017, \mn@doi [\mnras] {10.1093/mnras/stx1200}, \href
  {http://adsabs.harvard.edu/abs/2017MNRAS.470...20S} {470, 20}

\bibitem[\protect\citeauthoryear{{Saburova}, {Chilingarian}, {Katkov},
  {Egorov}, {Kasparova}, {Khoperskov}, {Uklein}  \& {Vozyakova}}{{Saburova}
  et~al.}{2018}]{Saburovaetal2018}
{Saburova} A.~S.,  {Chilingarian} I.~V.,  {Katkov} I.~Y.,  {Egorov} O.~V.,
  {Kasparova} A.~V.,  {Khoperskov} S.~A.,  {Uklein} R.~I.,   {Vozyakova} O.~V.,
   2018, \mn@doi [\mnras] {10.1093/mnras/sty2519}, \href
  {http://adsabs.harvard.edu/abs/2018MNRAS.481.3534S} {481, 3534}

\bibitem[\protect\citeauthoryear{{Salpeter}}{{Salpeter}}{1955}]{Salpeter1955}
{Salpeter} E.~E.,  1955, \mn@doi [\apj] {10.1086/145971}, \href
  {http://adsabs.harvard.edu/abs/1955ApJ...121..161S} {121, 161}

\bibitem[\protect\citeauthoryear{{Saulder}, {van Kampen}, {Chilingarian},
  {Mieske}  \& {Zeilinger}}{{Saulder} et~al.}{2016}]{Saulder2016}
{Saulder} C.,  {van Kampen} E.,  {Chilingarian} I.~V.,  {Mieske} S.,
  {Zeilinger} W.~W.,  2016, \mn@doi [\aap] {10.1051/0004-6361/201526711}, \href
  {http://adsabs.harvard.edu/abs/2016A%26A...596A..14S} {596, A14}

\bibitem[\protect\citeauthoryear{{Schlafly} \& {Finkbeiner}}{{Schlafly} \&
  {Finkbeiner}}{2011}]{Schlafly2011}
{Schlafly} E.~F.,  {Finkbeiner} D.~P.,  2011, \mn@doi [\apj]
  {10.1088/0004-637X/737/2/103}, \href
  {http://adsabs.harvard.edu/abs/2011ApJ...737..103S} {737, 103}

\bibitem[\protect\citeauthoryear{{Schmidt}, {J{\'o}zsa}, {Gentile}, {Oh},
  {Schuberth}, {Ben Bekhti}, {Winkel}  \& {Klein}}{{Schmidt}
  et~al.}{2014}]{Schmidt2014}
{Schmidt} P.,  {J{\'o}zsa} G.~I.~G.,  {Gentile} G.,  {Oh} S.-H.,  {Schuberth}
  Y.,  {Ben Bekhti} N.,  {Winkel} B.,   {Klein} U.,  2014, \mn@doi [\aap]
  {10.1051/0004-6361/201118170}, \href
  {http://adsabs.harvard.edu/abs/2014A%26A...561A..28S} {561, A28}

\bibitem[\protect\citeauthoryear{{Schombert}}{{Schombert}}{1998}]{Schombert1998}
{Schombert} J.,  1998, \mn@doi [\aj] {10.1086/300558}, \href
  {http://adsabs.harvard.edu/abs/1998AJ....116.1650S} {116, 1650}

\bibitem[\protect\citeauthoryear{{Seidel}, {Falc{\'o}n-Barroso},
  {Mart{\'{\i}}nez-Valpuesta}, {D{\'{\i}}az-Garc{\'{\i}}a}, {Laurikainen},
  {Salo}  \& {Knapen}}{{Seidel} et~al.}{2015}]{Seidel2015}
{Seidel} M.~K.,  {Falc{\'o}n-Barroso} J.,  {Mart{\'{\i}}nez-Valpuesta} I.,
  {D{\'{\i}}az-Garc{\'{\i}}a} S.,  {Laurikainen} E.,  {Salo} H.,   {Knapen}
  J.~H.,  2015, \mn@doi [\mnras] {10.1093/mnras/stv969}, \href
  {http://adsabs.harvard.edu/abs/2015MNRAS.451..936S} {451, 936}

\bibitem[\protect\citeauthoryear{{Sersic}}{{Sersic}}{1968}]{Sersic68}
{Sersic} J.~L.,  1968, {Atlas de galaxias australes}

\bibitem[\protect\citeauthoryear{{Shapiro}, {Gerssen}  \& {van der
  Marel}}{{Shapiro} et~al.}{2003}]{Shapiro2003}
{Shapiro} K.~L.,  {Gerssen} J.,   {van der Marel} R.~P.,  2003, \mn@doi [\aj]
  {10.1086/379306}, \href {http://adsabs.harvard.edu/abs/2003AJ....126.2707S}
  {126, 2707}

\bibitem[\protect\citeauthoryear{{Stewart}}{{Stewart}}{2017}]{Stewart2017}
{Stewart} K.~R.,  2017, in {Fox} A.,  {Dav{\'e}} R.,  eds,  Astrophysics and
  Space Science Library Vol. 430, Gas Accretion onto Galaxies. p.~249
  (\mn@eprint {arXiv} {1612.00513}), \mn@doi{10.1007/978-3-319-52512-9_11}

\bibitem[\protect\citeauthoryear{{Tody}}{{Tody}}{1986}]{iraf}
{Tody} D.,  1986, in {Crawford} D.~L.,  ed.,  Society of Photo-Optical
  Instrumentation Engineers (SPIE) Conference Series Vol. 627, Instrumentation
  in astronomy VI. p.~733

\bibitem[\protect\citeauthoryear{{Vollmer}, {Nehlig}  \& {Ibata}}{{Vollmer}
  et~al.}{2016}]{Vollmer2016}
{Vollmer} B.,  {Nehlig} F.,   {Ibata} R.,  2016, \mn@doi [\aap]
  {10.1051/0004-6361/201322899}, \href
  {http://adsabs.harvard.edu/abs/2016A%26A...586A..98V} {586, A98}

\bibitem[\protect\citeauthoryear{{Wang} et~al.,}{{Wang}
  et~al.}{2013}]{Wang2013}
{Wang} J.,  et~al., 2013, \mn@doi [\mnras] {10.1093/mnras/stt722}, \href
  {http://adsabs.harvard.edu/abs/2013MNRAS.433..270W} {433, 270}

\bibitem[\protect\citeauthoryear{{Wang} et~al.,}{{Wang}
  et~al.}{2015}]{Wang2015}
{Wang} J.,  et~al., 2015, \mn@doi [\mnras] {10.1093/mnras/stv1767}, \href
  {http://adsabs.harvard.edu/abs/2015MNRAS.453.2399W} {453, 2399}

\bibitem[\protect\citeauthoryear{{Wright} et~al.,}{{Wright}
  et~al.}{2010}]{Wright2010}
{Wright} E.~L.,  et~al., 2010, \mn@doi [\aj] {10.1088/0004-6256/140/6/1868},
  \href {http://adsabs.harvard.edu/abs/2010AJ....140.1868W} {140, 1868}

\bibitem[\protect\citeauthoryear{{Wyder} et~al.,}{{Wyder}
  et~al.}{2009}]{Wyder2009}
{Wyder} T.~K.,  et~al., 2009, \mn@doi [\apj] {10.1088/0004-637X/696/2/1834},
  \href {http://adsabs.harvard.edu/abs/2009ApJ...696.1834W} {696, 1834}

\bibitem[\protect\citeauthoryear{{Y{\i}ld{\i}z}, {Serra}, {Peletier},
  {Oosterloo}  \& {Duc}}{{Y{\i}ld{\i}z} et~al.}{2017}]{2017MNRAS.464..329Y}
{Y{\i}ld{\i}z} M.~K.,  {Serra} P.,  {Peletier} R.~F.,  {Oosterloo} T.~A.,
  {Duc} P.-A.,  2017, \mn@doi [\mnras] {10.1093/mnras/stw2294}, \href
  {https://ui.adsabs.harvard.edu/abs/2017MNRAS.464..329Y} {464, 329}

\bibitem[\protect\citeauthoryear{{Zabl} et~al.,}{{Zabl}
  et~al.}{2019}]{Zabl2019}
{Zabl} J.,  et~al., 2019, \mn@doi [\mnras] {10.1093/mnras/stz392}, \href
  {https://ui.adsabs.harvard.edu/abs/2019MNRAS.485.1961Z} {485, 1961}

\bibitem[\protect\citeauthoryear{{Zasov}, {Khoperskov}  \& {Tyurina}}{{Zasov}
  et~al.}{2004}]{Zasov2004}
{Zasov} A.~V.,  {Khoperskov} A.~V.,   {Tyurina} N.~V.,  2004, \mn@doi
  [Astronomy Letters] {10.1134/1.1795948}, \href
  {http://adsabs.harvard.edu/abs/2004AstL...30..593Z} {30, 593}

\bibitem[\protect\citeauthoryear{{Zasov}, {Khoperskov}  \& {Saburova}}{{Zasov}
  et~al.}{2011}]{Zasov2011}
{Zasov} A.~V.,  {Khoperskov} A.~V.,   {Saburova} A.~S.,  2011, \mn@doi
  [Astronomy Letters] {10.1134/S1063773711050069}, \href
  {http://adsabs.harvard.edu/abs/2011AstL...37..374Z} {37, 374}

\bibitem[\protect\citeauthoryear{{Zasov}, {Saburova}  \& {Abramova}}{{Zasov}
  et~al.}{2015}]{Zasovetal2015}
{Zasov} A.,  {Saburova} A.,   {Abramova} O.,  2015, \mn@doi [\aj]
  {10.1088/0004-6256/150/6/192}, \href
  {http://adsabs.harvard.edu/abs/2015AJ....150..192Z} {150, 192}

\bibitem[\protect\citeauthoryear{{Zhao}, {Arag{\'o}n-Salamanca}  \&
  {Conselice}}{{Zhao} et~al.}{2015}]{Zhao2015}
{Zhao} D.,  {Arag{\'o}n-Salamanca} A.,   {Conselice} C.~J.,  2015, \mn@doi
  [\mnras] {10.1093/mnras/stv190}, \href
  {https://ui.adsabs.harvard.edu/abs/2015MNRAS.448.2530Z} {448, 2530}

\bibitem[\protect\citeauthoryear{{Zhu} et~al.,}{{Zhu} et~al.}{2018}]{Zhu2018}
{Zhu} Q.,  et~al., 2018, \mn@doi [\mnras] {10.1093/mnrasl/sly111}, \href
  {http://adsabs.harvard.edu/abs/2018MNRAS.480L..18Z} {480, L18}

\bibitem[\protect\citeauthoryear{{de Blok} et~al.,}{{de Blok}
  et~al.}{2014}]{deBlok2014}
{de Blok} W.~J.~G.,  et~al., 2014, \mn@doi [\aap]
  {10.1051/0004-6361/201423880}, \href
  {http://adsabs.harvard.edu/abs/2014A%26A...569A..68D} {569, A68}

\bibitem[\protect\citeauthoryear{{den Heijer} et~al.,}{{den Heijer}
  et~al.}{2015}]{denHeijer2015}
{den Heijer} M.,  et~al., 2015, \mn@doi [Astronomische Nachrichten]
  {10.1002/asna.201412149}, \href
  {http://adsabs.harvard.edu/abs/2015AN....336..284D} {336, 284}

\bibitem[\protect\citeauthoryear{{van der Marel} \& {Franx}}{{van der Marel} \&
  {Franx}}{1993}]{vanderMarel1993}
{van der Marel} R.~P.,  {Franx} M.,  1993, \mn@doi [\apj] {10.1086/172534},
  \href {http://adsabs.harvard.edu/abs/1993ApJ...407..525V} {407, 525}

\makeatother
\end{thebibliography}

\label{lastpage}

\end{document}